\documentclass[reprint,superscriptaddress,nofootinbib,amsmath,amssymb,aps,floatfix]{revtex4-2}
\pdfoutput=1
\usepackage{graphicx}
\usepackage{dcolumn}
\usepackage{bm}
\usepackage[utf8]{inputenc}
\usepackage{graphics,graphicx}
\usepackage{parskip}
\usepackage{amsmath,amssymb}
\usepackage[normalem]{ulem}
\usepackage{amsfonts}
\DeclareMathOperator{\sign}{sign}
\usepackage{multirow}
\usepackage{fancyhdr}
\usepackage{xcolor}
\usepackage{placeins}
\usepackage[title]{appendix}
\usepackage{wasysym}
\usepackage{url}
\usepackage{braket}
\usepackage{float}
\usepackage[caption=false]{subfig}
\newcommand{\phantomsubfloat}[1]{
    {
        \captionsetup[subfigure]{labelformat=empty}
        \subfloat[][]{#1}
    }%
}
\usepackage{lipsum}
\usepackage{booktabs}
\usepackage{comment}
\usepackage{braket}
\usepackage[colorlinks=true,filecolor=.]{hyperref}
\usepackage{cleveref}
\usepackage{natbib}

\begin{document}

\title{Spectral properties of a three body atom-ion hybrid system}

\author{Daniel J. Bosworth} 
 \email{dan.bosworth@physnet.uni-hamburg.de}
\affiliation{%
 Zentrum f\"ur Optische Quantentechnologien, Universit\"at Hamburg,\\ Luruper Chaussee 149, 22761 Hamburg, Germany\\
}%
\affiliation{%
 The Hamburg Centre for Ultrafast Imaging, Universit\"at Hamburg,\\ Luruper Chaussee 149, 22761 Hamburg, Germany\\
}%

\author{Maxim Pyzh}%
\affiliation{%
 Zentrum f\"ur Optische Quantentechnologien, Universit\"at Hamburg,\\ Luruper Chaussee 149, 22761 Hamburg, Germany\\
}%

\author{Peter Schmelcher}
\affiliation{%
 Zentrum f\"ur Optische Quantentechnologien, Universit\"at Hamburg,\\ Luruper Chaussee 149, 22761 Hamburg, Germany\\
}%
\affiliation{%
 The Hamburg Centre for Ultrafast Imaging, Universit\"at Hamburg,\\ Luruper Chaussee 149, 22761 Hamburg, Germany\\
}%

\date{\today}

\begin{abstract}
We consider a hybrid atom-ion system consisting of a pair of bosons interacting with a single ion in a quasi-one-dimensional trapping geometry. Building upon a model potential for the atom-ion interaction developed in earlier theoretical works, we investigate the behaviour of the low-energy eigenstates for varying contact interaction strength $g$ among the atoms. In particular, we contrast the two cases of a static and a mobile ion. Our study is carried out by means of the Multi-Layer Multi-Configuration Time-Dependent Hartree method for Bosons, a numerically-exact \textit{ab initio} method for the efficient simulation of entangled mixtures. We find that repulsive atom interactions induce locally-distinct modifications of the atomic probability distribution unique to each eigenstate.
Whilst the atoms on average separate from each other with increasing $g$, they do not necessarily separate from the ion. The mobility of the ion leads in general to greater separations among the atoms as well as between the atoms and the ion. Notably, we observe an exchange between the kinetic energy of the atoms and the atom-ion interaction energy for all eigenstates, which is both interaction- and mobility-induced. For the ground state, we provide an intuitive description by constructing an effective Hamiltonian for each species, which aptly captures the response of the atoms to the ion's mobility. Furthermore, the effective picture predicts enhanced localisation of the ion, in agreement with our results from exact numerical simulations. 
\end{abstract}

\maketitle
\newpage

\section{Introduction}

Over the past decades, our understanding of the physics of neutral ultra-cold atom and laser-cooled ion systems has seen an unprecedented development, which has borne deep insights into their underlying and emergent physical phenomena. The superb degree of control achieved over these two quantum systems enables a high degree of accuracy and precision at both the single- and many-particle levels and has established them at the forefront of modern quantum many-body research. Recently, the two fields have been combined~\cite{smith2005cold,grier2009observation,schmid2010dynamics,zipkes2010trapped}, creating a versatile experimental platform for exploring fundamental interaction processes between atoms and ions at milli-Kelvin to micro-Kelvin temperatures~\cite{Harter2014cold,tomza2019cold}. The most prominent experimental and theoretical accomplishments to date include: studies on atom-ion collisions and reactions~\cite{ratschbacher2012controlling,hall2012millikelvin,harter2012single,meir2016dynamics,perezrios2020cold}
and related phenomena, such as the formation of chemical bonds~\cite{krukow2016reactive,krukow2016energy}, 
sympathetic cooling~\cite{zipkes2010trapped,zipkes2011kinetics,ravi2012cooling}
and charge transport~\cite{cote2000classical,Meinert2021Transport}; quantum simulation of condensed matter physics~\cite{bissbort2013emulating} and polaron models~\cite{Casteels2011polaronic,Jachymski2020Quantum}; quantum information investigations in the context of controlled entanglement generation~\cite{gerritsma2012bosonic,schurer2016impact} and decoherence effects~\cite{ratschbacher2013decoherence}; and precision measurements where the ion acts
as a local probe of the host gas' properties~\cite{schmid2010dynamics,zipkes2010trapped,veit2020pulsed}.\\ 
One of the on-going challenges faced by experimentalists in the field of atom-ion research is to create hybrid systems at nano-Kelvin temperatures. The nano-Kelvin scale marks the boundary of the ultra-cold regime, in which quantum phenomena dominate. The earliest hybrid traps were based on a straightforward superposition of optically-trapped atoms with ions confined in a Paul trap. The drawback of this scheme was found to be a heating mechanism caused by the excess micromotion of the ion~\cite{cetina2012micromotion,krych2015description}. In an effort to overcome this perceived limitation, several alternative schemes are currently being pioneered, such as: photoionisation of an atomic cloud using a femto-second laser~\cite{wessels2018absolute}, optical traps for ions~\cite{schneider2010optical,enderlein2012single,schaetz2017trapping,lambrecht2017long,schmidt2020optical}, and highly-excited Rydberg atoms within an atomic cloud~\cite{kleinbach2018ionic,schmid2018rydberg,Meinert2021Transport}.\\
In addition to the endeavours with alternative trapping schemes, proposals were also made to use the established Paul trap approach with a $^{6}\text{Li-}{}^{174}\text{Yb}^{+}$ hybrid mixture~\cite{tomza2015cold}, whose high mass-imbalance was predicted to undermine the micro-motion induced heating. This set-up was realised in a recent experimental breakthrough~\cite{feldker2020buffer}, reaching temperatures in the s- and p-wave scattering regime. In this regime of few partial waves, the increasingly-weighty quantum effects lead to deviations away from classical predictions and may enable the experimental observation of hitherto-unseen atom–ion Feshbach resonances~\cite{idziaszek2009quantum}. The presence of such a resonance would allow for complete control over the scattering parameters and thus, over the atom–ion interaction itself.\\
These recent experimental advances provide fresh impetus to extend the current theoretical understanding about the nature of the long-range atom-ion interaction at zero temperature. In this $T=0$ regime, the inelastic processes dominate the system's scattering dynamics and atoms can be captured in the weakly bound states of the atom-ion polarisation potential. These bound states enable the formation of so-called mesoscopic molecular ions, which are typically hundreds of nanometres in size~\cite{cote2002mesoscopic,schurer2017unraveling}. The amount of atoms captured by the ion is limited by the interaction strength among the atoms. The kinetic energy released during the capture process is distributed via phonon excitations among the unbound fraction and a density disturbance is created at the ion position~\cite{massignan2005static,schurer2015capture} or even, in the limit of a Tonks-Giradeau gas, a density bubble~\cite{goold2010ion}.\\
In our previous studies, we performed detailed analysis of the ground-state properties and dynamical behaviour of an atom–ion hybrid system for a static ion in a quasi-1D system~\cite{schurer2014ground,schurer2015capture}, analogous to the aforementioned $^{6}\text{Li-}{}^{174}\text{Yb}^{+}$ mixture. We have also examined the case of an equal mass system~\cite{schurer2017unraveling}, analogous to optically-trapped ions subject to the same external potential as the neutral atoms. In this work, we present an extension of both these investigations to the lowest-energy eigenstates of a few-body system composed of a single ion and two neutral bosonic atoms at zero temperature, with both species parabolically-confined in a quasi-1D geometry. The eigenstates of our few-body hybrid mixture are obtained via the Multi-Layer Multi-Configuration Time-Dependent Hartree method for Bosons (ML-MCTDHB)~\cite{Kr_nke_2013,CAO2013}, a numerically-exact \textit{ab initio} method that  efficiently accounts for the intra- and inter-species correlations via a time-dependent, variationally-optimised basis. Previously, ML-MCTDHB has been successfully applied to solve similar problems in mixtures of neutral bosonic  species~\cite{Keiler_2020,Theel_2020} and there is also an extension for dealing with mixtures of both bosonic and fermionic species~\cite{cao2017unified,chen2018entangle,kwasniok2020correl}. Our chosen numerical method requires interactions to be finite-valued at all spatial grid points of the system,
including at distances below the range of validity $R_0$ of the atom-ion interaction's long distance tail, which varies with the atom-ion separation $r$ as $-1/r^4$. To this end, we employ a model interaction potential whose parameters can be mapped to the real scattering parameters by means of quantum defect theory~\cite{schurer2014ground}. We characterise the five lowest-energy eigenstates of our few-body hybrid mixture across regimes of weak to strong atomic interactions through using the atom-atom and atom-ion distance correlations, number state composition and distribution of the total energy among the energy components.\\
This work is organised as follows. In~\cref{sec:model}, we introduce our model Hamiltonian, describe the form of the atom-ion interaction and present our numerical methodology. In~\cref{sec:results_spectrum}, we present the eigenstate spectrum of our three-body molecular ion system, before proceeding to examine the effects of varying interatomic interactions and ion mobility on the individual eigenstates in~\cref{sec:results_states}. In~\cref{sec:conclusion}, we conclude with a summary of our findings, examine the experimental viability of our model, and discuss prospective directions for future work.\\
\section{Atom-ion hybrid model and numerical approach}\label{sec:model}
In this section, we first present the Hamiltonian describing the atom-ion hybrid system in the laboratory frame (\cref{ssec:model}). We then introduce two alternative coordinate frames, which will prove themselves useful for the numerical
treatment and physical analysis (\cref{ssec:frames}). Finally, we provide a brief overview of the computational approach used throughout this work (\cref{ssec:mlx}) 
and define several physical quantities, which we will use to characterise 
the low-energy eigenstates (\cref{ssec:observs}). 

\subsection{Atom-ion hybrid model}\label{ssec:model}
We consider a system comprised of a single ion of mass $m_{I}$ and $N$ neutral bosonic atoms of mass $m_{A}$ at zero temperature.
We assume that both species are confined within quasi-1D parabolic traps, such that they can only move along the $z$-direction, with axial trapping frequency $\omega_{A}$ for the atoms and $\omega_{I}$ for the ion.
The atom-atom interactions are of s-wave character and are described by a contact pseudopotential.\\ 
When approaching a charged particle, the neutral atoms become polarised, resulting in long-range interactions between the ion and the induced dipole moments of the atoms. At large separations, the interaction between an atom at position $z_{A}$ and an ion at position $z_{I}$ behaves as $-\alpha e^2 / 2(z_{A} - z_{I})^4$, where $\alpha$ is the polarisability of the atom and $e$ is the elementary charge. This interaction introduces a new length $R^*=\sqrt{\alpha{e}^2m_{A} / \hbar^2}$ and energy scale $E^* = \hbar^2 / 2m_{A} {R^*}^2$ to the system, in addition to those set by the external traps. In atom-ion hybrid experiments~\cite{tomza2019cold}, the interaction range is typically $R^* \sim 100 \text{nm}$.\\
To properly account for interactions between the atomic and ionic species at all distances, whilst also ensuring our model is numerically tractable, we introduce a short-distance cut-off to the $1/r^4$ potential and describe the interaction at small separations by a repulsive barrier. The explicit model interaction used was developed previously in earlier works based on quantum defect theory~\cite{schurer2014ground,schurer2015capture} and can be expressed in units of $E^*$ and $R^*$ as
\begin{equation}
    \label{eq:atom-ion-int}
    V_{AI}(r) = v_0e^{-\gamma r^2} - \frac{1}{r^4 + \frac{1}{\kappa}},
\end{equation}
where $r = z_{A} - z_{I}$ denotes the atom-ion separation, $v_0$ the height and $\gamma$ the width of the repulsive short-range barrier, whilst $\kappa$ sets the short-range cut-off to the attractive tail and determines the number of bound states. 
It has been shown theoretically that at ultra-cold temperatures the rate of inelastic atom-ion collisions is larger for states with smaller binding energy~\cite{cote2002mesoscopic}. Accordingly, we choose our model parameters to be $v_0= 3\kappa$, $\gamma=4\sqrt{10\kappa}$ and $\kappa=80$, in units of $E^*$, ${R^*}^{-2}$ and ${R^*}^{-4}$, respectively. This choice accounts for the two uppermost bound states closest to the continuum $E = 0$~\cite{schurer2014ground}.

The species Hamiltonians $H_A$ and $H_I$ take the following form in units of $E^*$ and $R^*$: 
\begin{subequations}
    \begin{align}
        H_{A} & = \sum_{i=1}^N \bigg(     -\frac{\partial^2}{\partial {z^2_{Ai}}} + \frac{z^2_{Ai}}{l_A^4}\bigg) + \sum_{i<j}^N g\delta(z_{Ai} - z_{Aj}) \label{eq:h_atom}\\
        & = K_A+P_A+V_{AA}, \nonumber\\
        H_{I} & = - \beta \frac{\partial^2}{\partial z_I^2} + \frac{z_I^2}{l_A^4 \beta \eta^2} = K_I + P_I, \label{eq:h_ion}
    \end{align}
\end{subequations}
where $z_{Ai}$ denotes the position of the $i^{\text{th}}$ atom,$l_{A} = \sqrt{\hbar / m_{A}\omega_{A}} / R^*$ is the oscillator length of the parabolically-confined atoms re-scaled by $R^*$, $g$ is the effective strength of the atom-atom interaction, $\beta = m_{A}/m_{I}$ is the interspecies mass ratio, $z_I$ is the position of the ion and $\eta = \omega_{A}/\omega_{I}$ is the ratio of the trapping frequencies. $K_A$ and $K_I$ abbreviate the kinetic terms, $P_A$ and $P_I$ the external potentials and $V_{AA}$ the contact interaction. We fix $l_{A} =0.5$, $\eta=1$ and $N=2$ for the remainder of this work.\\
\begin{figure}
    \includegraphics[width = 0.75\columnwidth]{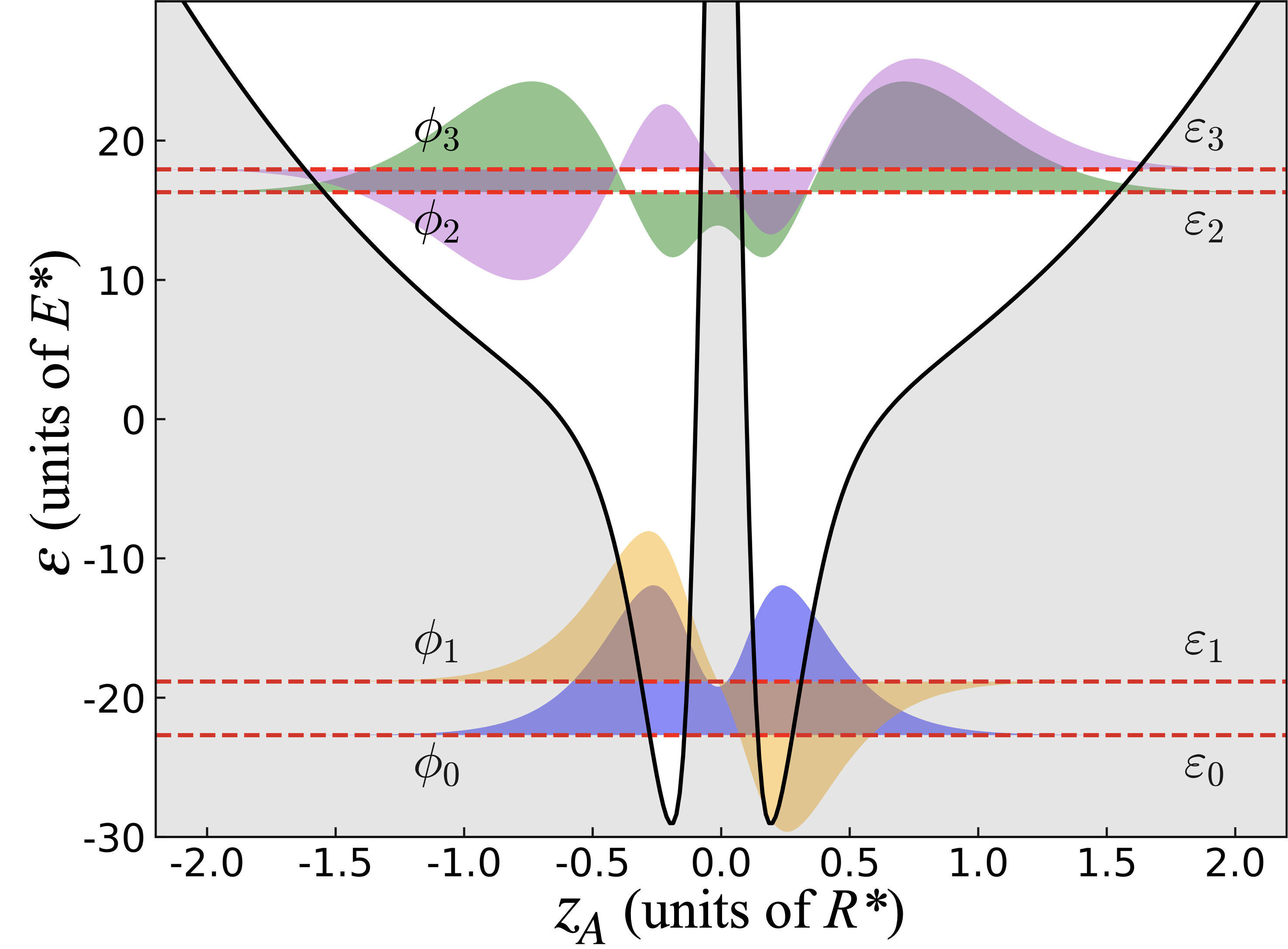}
    \caption{The first four lowest-energy solutions to the single-particle eigenvalue problem $h_{\text{1b}}\phi_i(z_{A}) = \varepsilon_i\phi_i(z_{A})$ (see eq.~\eqref{eq:h1b}), describing a single atom in a harmonic trap interacting with a static ion localised at $z=0$. The effective potential experienced by the atom is given by the solid black curve. The eigenstates $\{\phi_i(z_{A})\}_{i=0}^3$ (filled curves) are shown along red dashed lines, which indicate their eigenenergies $\{\varepsilon_i\}_{i=0}^3$. Energies and lengths are given in units of $E^*$ and $R^*$ set by the atom-ion interaction. The harmonic trap length is $l_{A} = 0.5$, in units of $R^*$.}
    \label{fig:h1b_solutions} 
\end{figure}
To provide some basic intuition for the system at hand, we assume that the ion is localised at the origin (i.e.: $l_I=0$, well-approximated by either a heavy ion or a tight trap) and that the atoms are non-interacting ($g=0$). In this case, the ion acts as a one-body potential for the atoms and does not receive any feedback from them. Our model reduces to a single-particle problem:
\begin{equation}\label{eq:h1b}
    h_{\text{1b}} = -\frac{\partial^2}{\partial z_{A}^2} + \frac{z_{A}^2}{{l_{A}}^4} + V_{AI}(z_{A}),
\end{equation}
which describes a single atom in an effective potential, being the superposition of the harmonic trap and atom-ion potential (see solid black curve in~\cref{fig:h1b_solutions}). The Schr\"odinger equation belonging to the one-body Hamiltonian $h_{\text{1b}}$ can be solved straightforwardly using exact diagonalisation. We choose to use a fast Fourier transform (FFT) discrete variable representation (DVR) basis~\footnote{Specifically, we use a FFT DVR basis of size $n=333$, which ensures the single particle eigenenergies are converged up to the sixth decimal place.}. The four lowest-energy single particle eigenstates $\{\phi_i(z)\}_{i=0}^3$ of the eigenvalue problem $h_{\text{1b}}\phi_i = \varepsilon_i\phi_i$ are depicted in~\cref{fig:h1b_solutions}. As mentioned above, our choice of parameters for the model interaction~\eqref{eq:atom-ion-int} results in two bound states in the atom-ion potential: $\phi_0$ and $\phi_1$. Due to the steep $-1/z^4$ contribution, these two bound states share a similar spatial extent ($\sim R^*$) and the peaks of their probability amplitudes coincide with the potential minima at $\approx\pm0.3 R^*$. In contrast, the higher-energy eigenstates $\phi_2$ and $\phi_3$ are extended across the harmonic trap ($\sim 2R^*$) with a significantly smaller probability amplitude at the potential minima. From now on, we will refer to $\phi_0$ and $\phi_1$ as {\it molecular} orbitals and $\phi_2$ and $\phi_3$ will be called {\it vibrational} orbitals.\\
In this work, we perturb this single-particle picture $h_{\text{1b}}$ two-fold: firstly, by considering interactions between the trapped pair of bosons and secondly, by including the motion of the ion. The former is parameterised by the contact interaction strength $g$ and the latter is parameterised by the mass ratio $\beta$, which determines the relative localisation between the two trapped species.\\
\subsection{Non-laboratory reference frames}\label{ssec:frames}
We emphasise that in the laboratory frame (LF), the atomic and ionic degrees of freedom are highly entangled because atoms can be bound to the mobile ion, which possesses a spatially-extended probability density. This fact makes it difficult to obtain well-converged numerical results. To account for these correlations, we replace the atom coordinates $z_{Ai}$ with the relative distances w.r.t.\ the ion $r_i = z_{Ai} - z_{I}$. The remaining ion coordinate $z_{I}$ can be retained (ion frame, IF) or replaced by the combined centre of mass $R=(m_I z_I+m_A \sum_i z_{Ai})/M$ of the system, where $M=m_I+N m_A$ is the system's total mass (centre of mass frame, CMF).\\
The primary frame used for the numerical simulations was the CMF, whose main advantage is its numerical stability during eigenstate acquisition and its more rapid convergence compared to the other frames. The corresponding CMF Hamiltonian is given by

\begin{equation}
    \label{eq:cm_ham_full}
    \begin{split}
        H = &\sum_{i=1}^N \Bigg( -\bigg(1+ {\beta}\bigg)\frac{\partial^2}{\partial r_i^2} + (1-d)\frac{r_i^2}{l_{A}^4} \Bigg) \\
        +&\sum_{i=1}^N \Bigg( v_0\exp{\big(-\gamma r_i^2\big) - \frac{1}{r_i^4 + \frac{1}{\kappa}}} \Bigg)\\
+&\sum_{i<j}\Bigg( g\delta(r_i - r_j) - 2\beta \frac{\partial}{\partial r_i}\frac{\partial}{\partial r_j} -\frac{2d}{l_{A}^4} r_i r_j\Bigg)\\
-&d\frac{\partial^2}{\partial R^2} + \frac{1}{l_{A}^4\beta\eta^2}(1 + N\beta\eta^2)R^2\\
+&\frac{2d}{l_{A}^4\beta\eta^2}(\eta^2 - 1)\sum_{i=1}^N Rr_i,
            \end{split}
\end{equation}
where the parameters have the same meanings as discussed in~\cref{ssec:model} and $d =\beta / (1 + N\beta)$. For equal trapping frequencies $\eta=1$, the above Hamiltonian decouples into two sub-Hamiltonians: one for the centre of mass co-ordinate $R$ and the other for relative co-ordinates $\{r_i\}$. The centre of mass sub-Hamiltonian $H_R(\eta=1) = -d\frac{\partial^2}{\partial R^2} + \frac{R^2}{l_{A}^4 d}$ describes a quantum harmonic oscillator of mass $M=1/2d$ and frequency $\Omega=2/l_{A}^2$ and can be solved analytically.
The atom-ion interaction now takes the form of a one-body potential, as was also the case for the static ion example discussed in~\cref{ssec:model}. However, the ion's motion induces two additional interactions between the relative coordinates, namely the positional ($r_i r_j$) and the derivative 
($\frac{\partial}{\partial r_i}\frac{\partial}{\partial r_j}$) couplings. In the limit of a static ion ($\beta\rightarrow0$), these additional interactions vanish and for $g=0$, we recover the one-body Hamiltonian~\eqref{eq:h1b} describing a single boson interacting with a static ion.\\
The other frame used for analysis was the IF. Whilst the IF is less efficient than the CMF, it is nonetheless more efficient than the LF since the entanglement between the interspecies degrees of freedom is reduced. 
Due to numerical instabilities however, it is challenging to obtain higher excited states in the IF, which
limits the analysis in this frame solely to the ground state.
Nevertheless, in contrast to the CMF, the IF provides access to the single particle atomic $\rho_1(z_{A})$ and ionic $\rho_1(z_{I})$ density distributions, which are laboratory frame quantities and thus allow for an easier interpretation (see Supplementary Material in~\cite{schurer2017unraveling}).
The IF Hamiltonian is given by
\begin{equation}
    \label{eq:ion_ham}
    \begin{split}
        H = &\sum_{i=1}^N \bigg( -(1+\beta)\frac{\partial^2}{\partial r_i^2} + \frac{r_i^2}{l_{A}^4} \bigg)\\ 
        +&\sum_{i=1}^N \bigg( v_0e^{-\gamma r^2} - \frac{1}{r^4 + \frac{1}{\kappa}} \bigg)\\
        -\beta&\frac{\partial^2}{\partial z_{I}^2} + \frac{1}{l_{A}^4}\bigg(N + \frac{1}{\beta\eta^2}\bigg)z_{I}^2\\
        +&\sum_{i<j} \bigg( g \delta(r_i - r_j) -2\beta\frac{\partial}{\partial r_i}\frac{\partial}{\partial r_j}\bigg)\\
        +2&\sum_{i=1}^N\bigg( \frac{z_{I} r_i}{l_{A}^4}  + \beta\frac{\partial}{\partial z_{I}}\frac{\partial}{\partial r_i}\bigg).
            \end{split}
\end{equation}
Note that the derivative coupling term ($\frac{\partial}{\partial r_i}\frac{\partial}{\partial r_j}$) is also present in this frame and that $z_{I}$ and $r_i$ cannot be decoupled for any choice of parameters.
\subsection{Computational approach}\label{ssec:mlx}
To solve for the lowest-energy eigenstates of our three-body problem, we employ the Multi-Layer Multi-Configuration Time-Dependent Hartree method for Bosons (ML-MCTDHB)~\cite{Kr_nke_2013,CAO2013}. ML-MCTDHB is a numerically-exact \textit{ab initio} method for performing time-dependent simulations of many-body quantum dynamics and it belongs to a wider family of multi-configuration Hartree-Fock methods~\cite{MEYER199073,BECK20001,Alon2008,KATO2004533}. In the same manner as its sibling methods, ML-MCTDHB utilises a variationally-optimised time-dependent basis which enables us to perform efficient calculations in a truncated Hilbert space, whilst ensuring that we fully cover the active sub-space of the complete Hilbert space. The multi-layer expansion allows for adopting the wavefunction ansatz to system-specific intra- and inter-species correlations. As a result, it is able to more efficiently treat mixtures with large numbers of particles in comparison to approaches that do not utilise multi-layering~\cite{ALON2007}.\\
The construction of the ML-MCTDHB wavefunction ansatz describing our three-body system proceeds as follows.
In the first step, we group together the indistinguishable degrees of freedom (DOF) and assign to them $S_{\sigma}\in\mathbb{N}$ species wave-functions 
$\{\ket{\psi_i^{\sigma}(t)}\}_{i=1}^{S_{\sigma}}$, 
with $\sigma$ denoting the distinct species. For our case, there are only two distinct species, corresponding to the atomic and ionic DOF.
Next, the total many-body wavefunction is written as a linear combination of product states:
\begin{equation}\label{eq:ansatz_top_layer}
    \ket{\psi(t)} = \sum_{i = 1}^{S_{I}}\sum_{j=1}^{S_{A}} A^{1}_{ij}(t) \ket{\psi_{i}^{\text{I}}(t)} \ket{\psi_{j}^{\text{A}}(t)},
\end{equation}
where $A^{1}_{ij}(t)$ are time-dependent top-layer coefficients and 
$\sigma=\text{A}$ stands either for $z_i$ or $r_i$, whilst $\sigma=\text{I}$ for $z_I$ or $R$, depending on the chosen frame. For $\eta=1$, the CMF DOF ($r_i$ and $R$) decouple, such that~\eqref{eq:ansatz_top_layer} becomes a single product state: $\ket{\psi(t)} = \ket{\psi^{\text{I}}(t)} \ket{\psi^{\text{A}}(t)}$. In such cases, the step in eq.\ \eqref{eq:ansatz_top_layer} is usually skipped as solving the sub-Hamiltonians independently using single-layer MCTDHB is more efficient.\\
In the second step, the species wavefunctions
$\ket{\psi_{i}^{\sigma}}$ for indistinguishable DOF are expanded in time-dependent number states $\ket{\textbf{n}}_t^{\sigma}$ 
to incorporate proper, in our case bosonic, quantum statistics:
\begin{equation}\label{eq:ansatz_2_layer}
    \ket{\psi_{i}^{(\sigma)}(t)} = \sum_{\textbf{n}|N_{\sigma}} A^{2;\sigma}_{i;\textbf{n}}(t) \ket{\textbf{n}}^{\sigma}_t,
\end{equation}
with time-dependent species-layer coefficients 
$A^{2;\sigma}_{i;\textbf{n}}(t)$.
The number states 
$\ket{\textbf{n}}^{\sigma}=(n_1, \ldots ,n_{s_{\sigma}})$ 
are comprised of $s_{\sigma} \in \mathbb{N}$ 
time-dependent single-particle functions (SPFs) 
$\{\ket{\phi_i^{\sigma}(t)}\}_{i=1}^{s_{\sigma}}$ 
and the sum goes over all possible number state configurations $\textbf{n}|N_{\sigma}$ which
fulfil the constraint of a fixed number of particles
$\sum_{i=1}^{s_{\sigma}} n_i=N_{\sigma}$ .\\
Finally, the time-dependent SPFs are represented on a one-dimensional discrete variable representation (DVR) basis $\{\ket{\chi^{\sigma}_i}\}_{i=1}^{\mathcal{M}_{\sigma}}$ (in our case, a FFT DVR basis)
\begin{equation}\label{eq:ansatz_3_layer}
    \ket{\phi_i^{\sigma}(t)} = \sum_{j=1}^{\mathcal{M}_{\sigma}} A^{3;\sigma}_{ij}(t) \ket{\chi^{\sigma}_j},
\end{equation}
with time-dependent particle-layer coefficients $A^{3;\sigma}_{i;j}(t)$~\cite{Light1985}.\\
The Hilbert space of our system is truncated at each layer and controlled through the values of $S_{\sigma}$, $s_{\sigma}$ and $\mathcal{M}_{\sigma}$. 
This allows us to tailor our ansatz to suit the degree of intra- and inter-species correlations present in the system. Note that in contrast to standard approaches, the SPFs are time-dependent, allowing for a considerable boost in computation time.\\
The equations of motion for the three layers of coefficients $A^{1}_{ij}$, $A^{2;\sigma}_{i;\textbf{n}}$ and $A^{3;\sigma}_{i;j}$ outlined above are derived from the Dirac-Frenkel variational principle~\cite{Beck2000TheWavepackets}:
\begin{equation}\label{eq:dirac_frenkel}
    \bra{\delta\psi}(i\partial_t - \hat{H})\ket{\psi} = 0.
\end{equation}
The ground state and higher excited states are obtained by means of improved relaxation of an initial input state. Specifically, ML-MCTDHB propagates the non-top layer coefficients $A^{2;\sigma}_{i;\textbf{n}}$ and $A^{3;\sigma}_{i;j}$ in imaginary time to a fixed point, at which point it diagonalises the top-layer $A^{1}_{ij}$ matrix. This process is performed recursively until the top-layer and non-top layer coefficients become constant during the diagonalisation and imaginary-time propagation, respectively. This converged result is a stationary state of the system, which lies in the truncated Hilbert space given by the ML-MCTDHB ansatz. Different stationary states can be obtained by carefully selecting the initial input wave function provided to the improved relaxation routine.\\
Working in the CMF and IF offers a distinct numerical advantage 
to working in the LF since, as discussed in~\cref{ssec:frames},
position correlations between the atom and ion in a bound pair are accounted for implicitly within the relative coordinate 
$r_i = z_{Ai} - z_{I}$. 
This enables us to further truncate our active Hilbert space on the top-layer~\eqref{eq:ansatz_top_layer} in these frames, resulting in greater computational efficiency. We emphasise that this additional level of truncation is only possible due to the multi-layer structure of our wavefunction ansatz.\\
\subsection{Observables}\label{ssec:observs}
Here we introduce several physical quantities used to characterise the eigenstates in~\cref{sec:results_states}.
\subsubsection{Inter-atomic and inter-species separation distributions}\label{ssec:sep_dists}
In the decoupled CMF ($\eta=1$), we focus on the relative sub-Hamiltonian 
(see eq.\ \eqref{eq:cm_ham_full}) with bosonic DOF $r_i$. 
For a system of two atoms with relative positions $r$ and $r^{\prime}$ to the ion, the wavefunction takes the form $\psi(r,r^{\prime})$ and the probability density for finding the atoms in the configuration $(r,r^{\prime})$ is given by
the density $\rho_2(r,r^{\prime}) = \psi(r,r^{\prime})^*\psi(r,r^{\prime})$.\\
We are now able to extract a useful quantity related to LF coordinates, 
namely the interatomic separation distribution 
$\rho_1(z_{A} - z_{A}^{\prime})$. 
To this end, we note that 
\begin{equation}
    \int dr dr'\; \rho_2(r,r') = \int dX dY\; \tilde{\rho}_2(X,Y)=1,
\end{equation}
where in the second step we do a coordinate transformation 
$X = r - r^{\prime}$, $Y = (r + r^{\prime})/2$ and $\tilde{\rho}_2(X,Y)=\rho_2(r(X,Y),r'(X,Y))$.
Now by integrating out the coordinate $Y$, we obtain the reduced one-body density $\rho^{AA}_1(X)$:
\begin{equation}\label{eq:rho1r}
    \rho^{AA}_1(X = z_A-z'_A) = \int dY\; \tilde{\rho}_2(X,Y).
\end{equation}
Additionally, in both the CMF and IF we can evaluate the expectation values 
for the atom-atom and atom-ion separations:
\begin{equation}\label{eq:d_AA}
 \braket{d_{AA}}=\int dX\; |X| \rho^{AA}_1(X),
\end{equation}
\begin{equation}\label{eq:d_AI}
\braket{d_{AI}}=\int dr\; |r| \rho_1(r),
\end{equation}
where in the IF $\rho_1(r)= \int dz_I dr'\; \psi^*(z_I,r,r')\psi(z_I,r,r')$.
	 Whereas in the CMF, $\rho_1(r)= \int dr'\; \rho_2(r, r')$ is the one-body probability density of the relative coordinate $r=z_A-z_I$, giving us the interspecies separation distribution.\\
\subsubsection{Bunching probability}
In this paper, we often refer to the atoms as being \textit{'bunched'} or \textit{'anti-bunched'}. To clarify what is meant by this quantitatively, we define the so-called bunching probability as the total probability for the atoms to be found on the same side of the ion, irrespective of their separation. This can be written explicitly as follows:
\begin{equation}
    \begin{split}
        P_\text{bunched} = &\int_{-\infty}^{0} \int_{-\infty}^{0} dr dr^{\prime} \rho_2(r,r^{\prime})\\ 
        &+ \int_{0}^{\infty} \int_{0}^{\infty} dr dr^{\prime} \rho_2(r,r^{\prime}),
    \end{split}
\end{equation}
i.e. the probability to be found in the lower-left or upper-right quadrants of the two-particle density $\rho_2(r,r^{\prime})$. 
Naturally, $P_{\text{anti-bunched}} = 1 - P_\text{bunched}$.
In the bunched configuration, atoms favour the same side of the ion 
$P_\text{bunched}>P_{\text{anti-bunched}}$, whereas 
in the anti-bunched configuration, the atoms 
are more likely to be found on opposite sides of the ion
$P_\text{bunched}<P_{\text{anti-bunched}}$.\\
\section{Low-Energy Spectrum}\label{sec:results_spectrum}
In this section, we analyse how the five lowest-lying eigenenergies of our hybrid model (see~\cref{sec:model}) change under variation of the inter-atomic interaction strength $g$
and elaborate on the differences between a static and a mobile ion.\\
\begin{figure}
    \centering
    \includegraphics[width=0.6\columnwidth]{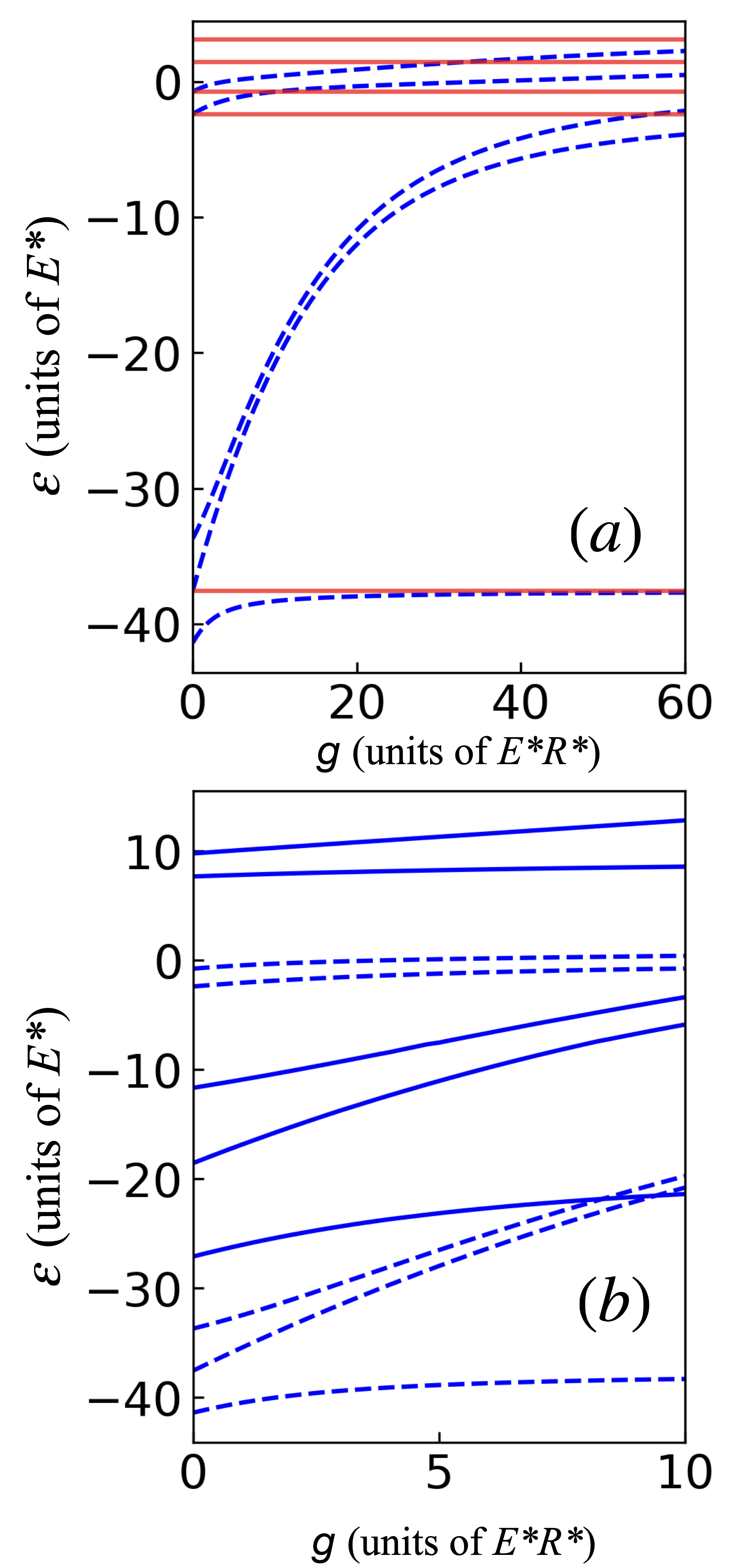}
    \phantomsubfloat{\label{fig:static_spec_ext}}
    \phantomsubfloat{\label{fig:spectrum}}
    \vspace{-2\baselineskip}
    \caption{
    Low-energy spectrum of two bosons coupled to a single ion as a function of the atom-atom contact interaction strength $g$ for (a) a static ion $\beta=0$ (dashed lines) and (b) a mobile ion $\beta=1$ (solid lines). The corresponding energies for two non-interacting fermions are indicated in (a) by the horizontal solid red lines. Note that the eigenenergies should approach the fermionic values in the Tonks-Girardeau limit $g\rightarrow\infty$. The static ion spectrum from (a) is given additionally for reference in (b) (dashed lines). Note the different range of $g$ values in the subfigures (a) and (b).
    }
    \label{fig:spectra}
\end{figure}
We first discuss the spectrum for the case of a static ion pinned at $z_{I}=0$, 
where the atom-ion interaction (see eq.~\eqref{eq:atom-ion-int})
reduces to an effective one-body potential. 
The first five eigenenergies are given by the blue dashed lines in~\cref{fig:static_spec_ext}. 
They increase monotonically with $g$ and approach the Tonks-Girardeau (TG) energies  ($g\rightarrow\infty$), equivalent to those of two non-interacting fermions 
subject to the same one-body potential (solid red lines)~\cite{Girardeau2006BosonizationGas}.
The ground state at $g=0$ corresponds to the bosonic number state $\ket{2,0,0,0}$ 
built from SPFs of $h_{1b}$ (eq.\ \eqref{eq:h1b}) (see also~\cref{fig:h1b_solutions}). 
It saturates rapidly to the corresponding TG energy 
of the fermionic number state $\ket{1,1,0,0}$, tapering off beyond $g=4$.\\
The first and second excited states at $g=0$ correspond to 
excitations of one or both atoms to the second molecular orbital, i.e.
$\ket{1,1,0,0}$ and $\ket{0,2,0,0}$. We observe that with increasing $g$, the energy gap between these states, given by
$\epsilon_2-\epsilon_1$, first decreases
up to $g = 10$ before increasing again and then tapering off at large $g$ as the system approaches the TG limit ($g\rightarrow\infty$). In this limit, the interacting bosons which comprise the first and second excited state are energetically-mapped to pairs of non-interacting fermions with number state configurations $\ket{1,0,1,0}$ and $\ket{1,0,0,1}$, respectively. The energy gap between these states is equal to the gap at $g=0$ between the third and fourth excited states $\epsilon_4-\epsilon_3$.\\
The third and fourth excited states at $g=0$ correspond to excitations of a single atom to 
one of the vibrational orbitals, i.e. $\ket{1,0,1,0}$ and $\ket{1,0,0,1}$. 
They are quite robust to $g$ variation, 
being a consequence of the reduced spatial overlap between the molecular and vibrational orbitals.
In the TG limit, they map to the fermionic states $\ket{0,1,1,0}$ and $\ket{0,1,0,1}$ 
and as a result, they have the same energy gap as at $g=0$.\\
The ion's mobility has two effects on the spectrum (blue solid lines in~\cref{fig:spectrum}): 
(i) a positive energy shift for all states and 
(ii) increased energy separation among the eigenstates. Aside from this, we still observe a monotonous increase of the energies with $g$. Interestingly, we also observe a tapering off of the energy of the ground state at large $g$, which is reminiscent of the energy mapping between hard-core bosons and non-interacting fermions. Formally however, the criteria for the TG mapping are not fulfilled since firstly, we do not have a single- but rather a two-component system and secondly, the hard-core interaction exists only between the atoms.
\section{Analysis of Eigenstates}\label{sec:results_states}
In this section, we will examine in detail the individual eigenstates comprising the low-energy spectrum presented in~\cref{sec:results_spectrum}, from the ground state up to the fourth excited state (\cref{ssec:s0,ssec:s1,ssec:s2,ssec:s3,ssec:s4}). In particular, we will explore the effect of varying atomic interactions and ion mobility on the properties of the eigenstates. 
For each state considered, we will analyse the distribution of energy among the various energy components and discuss what implications this has for the inter-atomic and inter-species separation distributions introduced in~\cref{ssec:sep_dists}. Moreover, we will also consider to what extent the single-particle picture $h_{1b}$ based on eq.\ \eqref{eq:h1b} is modified by exploring the number state composition of each eigenstate. Each sub-section focuses on a specific eigenstate and begins with a short summary of the main physical properties of that state.
\subsection{Ground state}\label{ssec:s0}
In the ground state, both atoms bind to the ion in the lowest bound-state and show no preference for bunching or anti-bunching when they are non-interacting and the ion is static. Finite interactions between the atoms cause them to separate to opposite sides of the ion (see~\cref{sssec:s0_static}). 
For an equal mass system ($\beta=1$), the ion’s mobility results in a slight preference for the non-interacting atoms to be bunched, which can be understood using an effective potential model that shows the atom pair clusters at the trap centre when the ion is mobile (see~\cref{sssec:s0_mobile_atom}). This interspecies correlation effect competes against the interatomic anti-correlations, delaying the on-set of complete separation of the atoms. The fully-separated atom pair pins the mobile ion from either side, such that it becomes increasingly localised at the trap-centre (see~\cref{sssec:s0_mobile_ion}).
\begin{figure*}
    \centering
    \includegraphics[width=1.9\columnwidth]{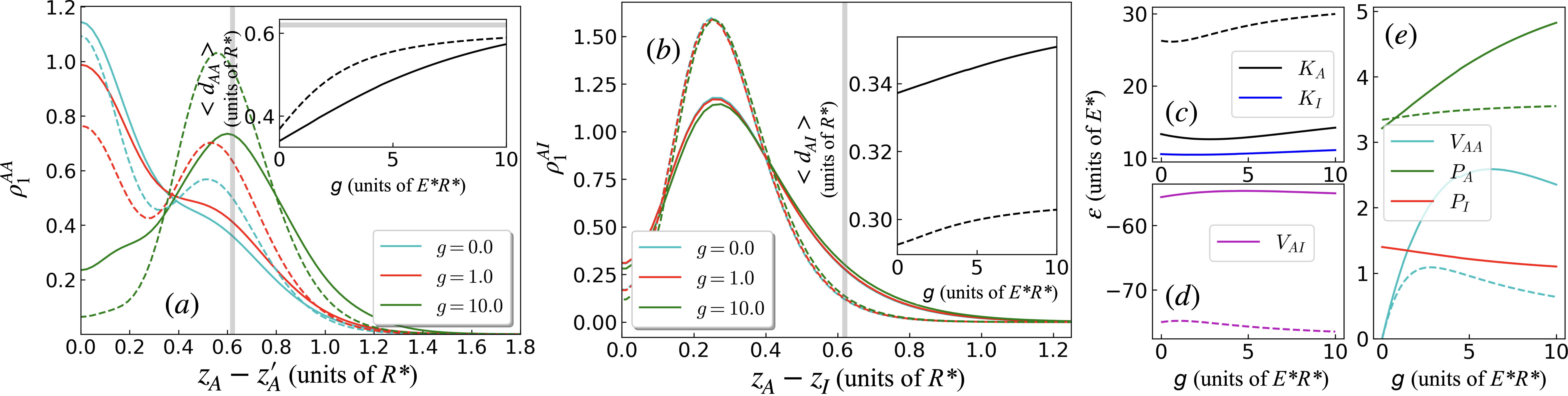}
    \phantomsubfloat{\label{fig:sep_s0}}
    \phantomsubfloat{\label{fig:AI_sep_s0}}
    \phantomsubfloat{\label{fig:ecpt_s0}}
    \phantomsubfloat{\label{fig:ecpt_s0_2}}
    \phantomsubfloat{\label{fig:ecpt_s0_3}}
    \vspace{-2\baselineskip}
    \caption{Key observables for the ground state. 
        \textbf{(a):} interatomic separation distribution $\rho^{AA}_1(z_{A} - z_{A}^{\prime})$ for different atom-atom coupling strengths $g$. The inset shows the expectation value $\braket{d_{AA}}$ for the atom-atom separation as a function of $g$ (eq.~\eqref{eq:d_AA}).  \textbf{(b):} interspecies separation distribution $\rho^{AI}_1(z_{A} - z_{I})$ for varying atom-atom coupling strengths $g$. The inset shows the expectation values $\braket{d_{AI}}$ for the atom-ion separation as a function of $g$ (eq.~\eqref{eq:d_AI}). \textbf{(c)-(e):} the evolution of the laboratory frame energy components with atom-atom coupling strength $g$. \textit{Note for (a) and (b):} the grey lines indicate the distance between the minima of the atom-ion interaction potential \eqref{eq:atom-ion-int}. Due to the parity symmetry it is sufficient to show only the positive semi-axis. \textit{All subfigures:} the solid curves correspond to a mobile ion, whilst dashed curves correspond to a static ion.
    }
     \label{fig:gs_observables}   
\end{figure*}

\begin{figure*}
    \centering
    \includegraphics[width=1.8\columnwidth]{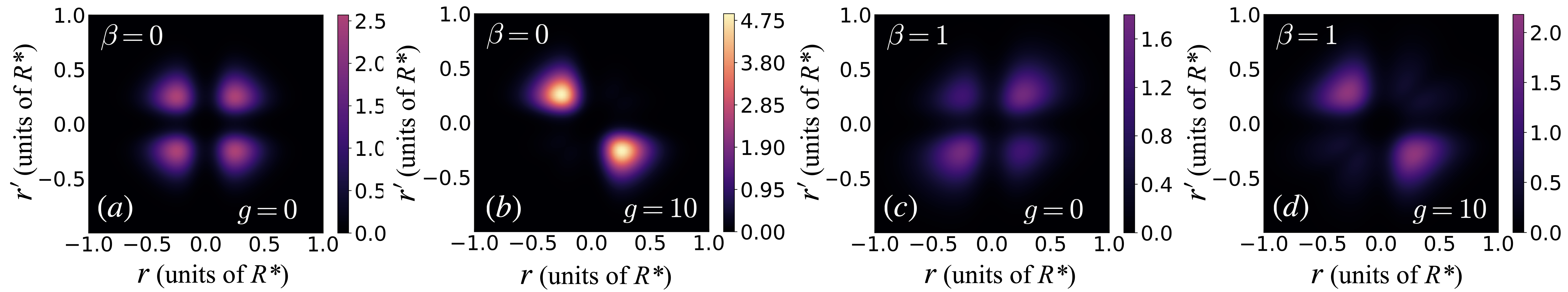}
    \phantomsubfloat{\label{fig:dmat2_s0_1}}
    \phantomsubfloat{\label{fig:dmat2_s0_2}}
    \phantomsubfloat{\label{fig:dmat2_s0_3}}
    \phantomsubfloat{\label{fig:dmat2_s0_4}}
    \vspace{-2\baselineskip}
    \caption{
        Snapshots of the atomic probability density $\rho_2^{\text{AA}} = |\psi(r,r^{\prime})|^2$ of the ground state for different interaction strengths $g$ for a static ion ((a) and (b)) and a mobile ion ((c) and (d)).
    }
    \label{fig:dmat2_s0}
\end{figure*}

\subsubsection{Static ion}\label{sssec:s0_static}
The ground state of two non-interacting ($g=0$) atoms 
coupled to a static ($\beta=0$) ion located at the trap centre $z_{I} = 0$
is given by the number state $\ket{2,0,0,0}$ 
w.r.t.\ the single-particle eigenstates of $h_{\text{1b}}$ (see eq.~\eqref{eq:h1b}), i.e.,
both atoms occupy the lowest molecular orbital $\phi_0$ in~\cref{fig:h1b_solutions}. 
The bunched and anti-bunched configurations are equally probable 
(see ~\cref{fig:dmat2_s0_1}), 
indicated also by the two peaks in the interatomic separation distribution 
$\rho_1^{AA}(z_A - z_A^{\prime})$ 
(light-blue dashed curve in~\cref{fig:sep_s0}).\\
With increasing $g$, 
we observe in~\cref{fig:sep_s0} 
a depletion of the central peak at $z_{A} = z_{A}^{\prime}$  
in favour of the side humps, 
which smoothly shift their position to larger separations. As a result, the atom-atom separation $d_{\text{AA}}$ increases 
(black dashed line in the inset of~\cref{fig:sep_s0}). The sharp initial growth in the total energy (see~\cref{fig:spectrum}) 
can be mainly attributed
to the behaviour of the intra-atomic interaction $V_{\text{AA}}$, which increases monotonously for $g<3$ and decreases thereafter as the probability for the atoms to occupy the same position gradually vanishes
(light-blue dashed curve in~\cref{fig:ecpt_s0_3}). In addition, there is a near 1:1 exchange between the atomic kinetic $K_A$
and the atom-ion interaction $V_{AI}$ energies with increasing interaction strength $g$: the increased probability for the anti-bunched configuration enables the atoms to localise more around the atom-ion potential minimum (increasing $K_{A}$, black dashed line in~\cref{fig:ecpt_s0}) and slide down within the $V_{AI}$ potential (decreasing $V_{\text{AI}}$, pink dashed line in~\cref{fig:ecpt_s0_2}).
Thus the distance between the atoms and the ion $d_{AI}$
(black dashed line in the inset of~\cref{fig:AI_sep_s0})
is almost unchanged, though it shows a gradual increase due to the increasingly depleted $z_{A}=z_{I}$ region (see dashed curves in~\cref{fig:AI_sep_s0}) which is further reflected in the gentle increase of the atomic trap potential energy $P_{A}$ (green dashed line in~\cref{fig:ecpt_s0_3}) since the atoms have a reduced probability to be found at the origin $z_{A} = 0$.\\
At stronger interactions ($g=10$), the bunched configuration becomes
completely suppressed and the atoms 
exist on opposite sides of the ion (see ~\cref{fig:dmat2_s0_2}). 
While approaching the TG regime, all energies begin to saturate 
and $d_{\text{AA}}$ approaches the corresponding limit
of the separation between two non-interacting fermions $\approx0.6$, i.e.\ the distance between the $V_{AI}$ minima (see~\cref{fig:h1b_solutions}). $V_{AA}$ decreases for large $g$ and will approach zero in the TG limit.
\subsubsection{Mobile ion: impact on the atoms}\label{sssec:s0_mobile_atom}
\begin{figure}
    \centering
    \includegraphics[width=0.75\columnwidth]{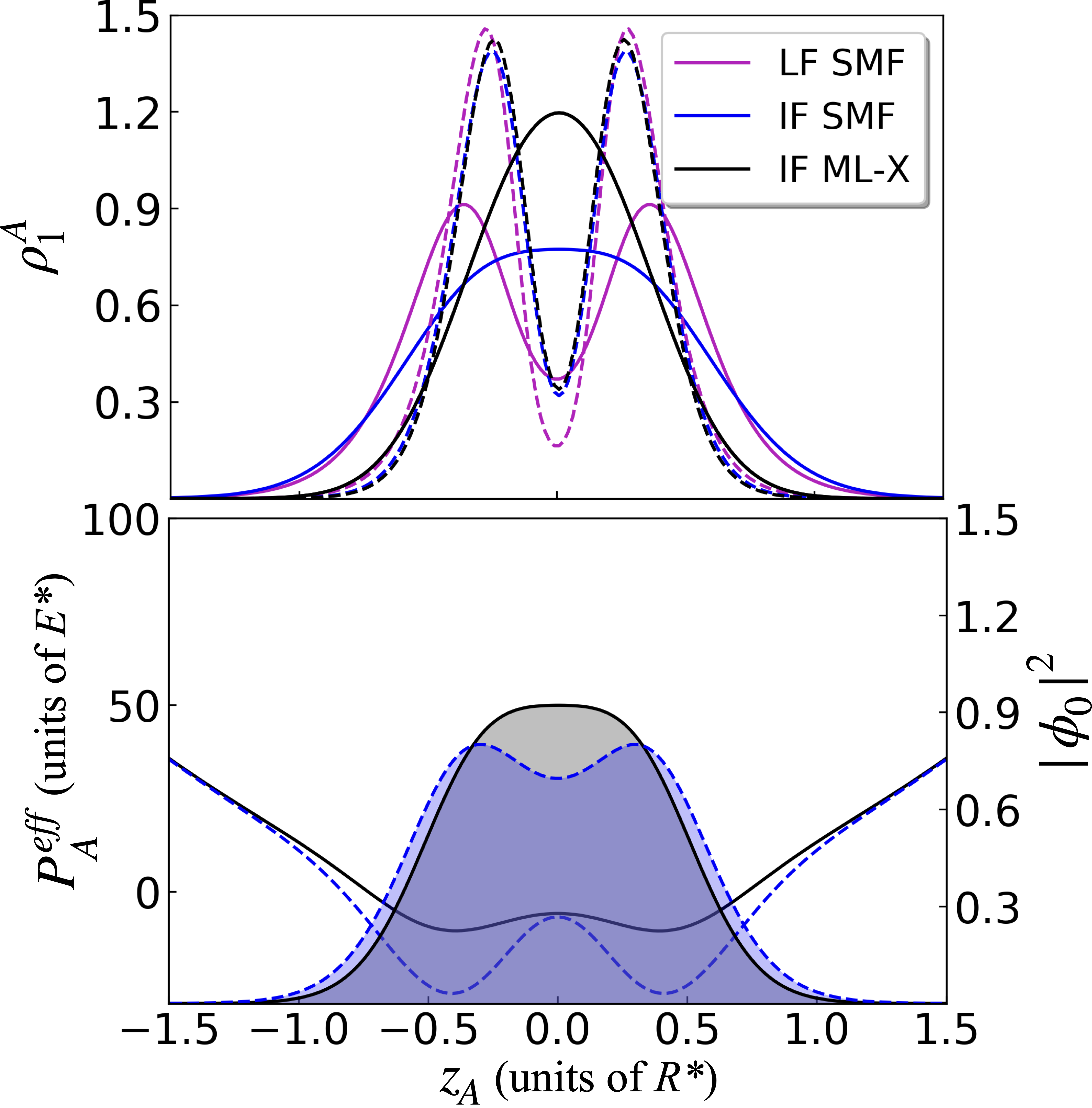}
    \phantomsubfloat{\label{fig:pA}}
    \phantomsubfloat{\label{fig:eff_pot_atom}}
    \vspace{-2\baselineskip}
    \caption{
    (a) atomic density $\rho_1(z_{A})$ for the case of a heavy ion ($\beta=0.034$, dashed lines) and a mobile ion ($\beta=1$, solid lines), obtained via: the lab frame species mean field (LF SMF) ansatz (pink curves), the ion frame SMF (IF SMF) ansatz (blue curves) and the exact IF result from the full ML-MCTDH ansatz (see eq.~\eqref{eq:ansatz_top_layer}) (IF ML-X) (black curves). (b) effective potential $P_A^{eff}(z_{A})$ experienced by the atoms due to the IF ML-X density $\rho^{IF}_1(z_{I})$ (solid black line) and that obtained with the IF SMF ansatz $\tilde{\rho}^{IF}_1(z_{I})$ (dashed blue line), derived from eq.~\eqref{eq:eff_ham} with $\beta=1$. The filled curves denote the respective ground state orbitals $|\phi_0|^2$ calculated from the potential. Note for both subfigures that the atoms are non-interacting $g=0$.
    }
\end{figure}
Let us now consider the impact of the ion's mobility ($\beta=1$) on the ground state properties.
The two-body density $\rho_2(r,r')$ becomes more spread out
(compare~\cref{fig:dmat2_s0_1,fig:dmat2_s0_3}), 
which results in a decrease of the kinetic energy of the atoms $K_A$ and a positive shift 
in the atom-ion interaction energy $V_{AI}$
(see black and pink solid curves in~\cref{fig:ecpt_s0,fig:ecpt_s0_2}). Similarly, the atom-ion separation distribution $\rho_1^{AI}(r)$ broadens, leading to an increase in the atom-ion separation $d_{AI}$ (compare solid and dashed lines in inset of~\cref{fig:AI_sep_s0}).
By comparing the interatomic separation distributions $\rho_1^{AA}$
between the static and mobile cases (dashed and solid lines in~\cref{fig:sep_s0}),
we infer that the additional derivative and positional coupling terms in eq.~\eqref{eq:cm_ham_full},
introduced by the ion mobility, impede the process of particle separation.
Thus when $\beta=1$, the average atom-atom distance $d_{AA}$ 
reaches values of the static ion system only at a stronger coupling $g$
(compare black curves in the inset of~\cref{fig:sep_s0}). \\
To obtain a better understanding of this mobility-induced bunching effect (see~\cref{fig:dmat2_s0_3}), 
we now examine the ground state through the lens of species mean-field (SMF) theory.
This will allow us to extract for each species an effective one-body potential 
induced by the other component, 
effectively decoupling the equations of motion between the distinguishable DOF.
As already discussed in~\cref{ssec:frames}, the LF is badly-suited for this purpose.
On the other hand, the IF incorporates 
the correlations of a bound atom-pair following the ion movement
and moreover, allows us to obtain several useful physical quantities of the LF.
In the notation of ML-MCTDHB, the IF SMF ansatz assumes a single product state on the top layer \eqref{eq:ansatz_top_layer}:
\begin{equation}\label{eq:product_ansatz}
    \psi\big(z_{I},r_1,\ldots,r_N) \approx \psi_{A}\big( r_1,\ldots,r_N \big) \psi_{I}\big(z_{I} \big).
\end{equation}
In~\cref{fig:pA}, we compare
the one-body density $\rho_1(z_{A})$ of the atoms 
obtained with the SMF ansatz (blue lines) to that of the exact ML-MCTDHB solution (black lines) in the IF for non-interacting atoms $g=0$ bound to a mobile ($\beta =1$) as well as a heavy, near-static ($\beta=0.034$) ion. The latter mass ratio corresponds to the species pairing $^{6}\text{Li-}{}^{174}\text{Yb}^{+}$. We additionally show the results obtained via the SMF ansatz in the LF (pink lines) for comparison. For a heavy ion, 
the SMF ansatz is well justified in both frames
(compare dashed curves in~\cref{fig:pA}). 
The atoms are most likely to be found around the minima of the atom-ion potential at $\approx\pm0.3 R^*$.
For a mobile ion ($\beta=1$), the exact result shows that the atoms are now most likely to be found at the trap centre (see solid black line in~\cref{fig:pA}). Whilst the IF SMF ansatz approximately captures this feature, it nonetheless shows substantial quantitative deviations from the exact result. Thus the entanglement between the DOF $z_I$ and $r_i$, neglected in the IF SMF, 
favours increased bunching of atoms around the centre of the harmonic trap. The LF SMF ansatz, which neglects entanglement between the DOF $z_I$ and $z_{A}^i$, still predicts that the atoms are most likely to be found around the minima of the static ion potential. The entanglement between the DOF $z_I$ and $z_{A}^i$ is therefore crucial for capturing the correct form of the probability density.\\
In the following, we ignore the entanglement between $z_I$ and $r_i$ and aim at understanding the qualitative features of $\rho_1(z_A)$ for a mobile ion ($\beta=1$) with non-interacting atoms ($g=0$).
To this end, we perturb atomic Hamiltonian in the LF~\eqref{eq:h_atom} with an effective atom-ion interaction potential found by integrating out the ionic degree of freedom in the interspecies interaction~\eqref{eq:atom-ion-int}. The result is:
\begin{equation}\label{eq:eff_ham}
\begin{split}
H_{A}^{eff} &= K_A + P_A + V_{AA} + \sum_{i=1}^N \int dz_{I} \; 
V_{\text{AI}}(z_{A}^i,z_{I}) \tilde{\rho}^{IF}_1(z_{I})\\ &= K_A + P_A^{eff} + V_{AA},
\end{split}
\end{equation}
where $\tilde{\rho}^{IF}_1(z_{I})$ is the approximate one-body density of the ion 
obtained with the IF SMF ansatz.
Note, $\tilde{\rho}^{IF}_1(z_{I})$ is different to the density $\tilde{\rho}^{LF}_1(z_{I})$ 
one would normally use based on the LF SMF ansatz.
Since $\tilde{\rho}^{IF}_1(z_{I})$ incorporates some of the many-body correlations between laboratory DOF,
we expect this effective Hamiltonian to better capture 
the behaviour of the atomic species than $\tilde{\rho}^{LF}_1(z_{I})$.
Indeed, we observe for a mobile ion ($\beta=1$) that the extended ion density flattens out the $V_{AI}$ minima and central barrier,
yielding an effective potential which takes the form of a harmonic potential with a small modulation around the origin (blue dashed lines in~\cref{fig:eff_pot_atom}). The corresponding ground state orbital $|\phi_0|^2$ (filled dashed blue curve in~\cref{fig:eff_pot_atom}) shows increased probability for the atom to be found at the centre of the harmonic trap, though it still displays peaks around the remnants of potential minima. The effective potential obtained from eq.~\eqref{eq:eff_ham} with the exact ion density $\rho^{IF}_1(z_{I})$ is qualitatively similar, however the modulation around the origin is considerably weaker and therefore its ground state orbital $|\phi_0|^2$ bears closer resemblance to a Gaussian (solid black lines in~\cref{fig:eff_pot_atom}), in accordance with the shape observed in~\cref{fig:pA}. Whilst the effective picture obtained using the IF SMF ansatz provides a qualitatively correct description of the mobility-induced atomic bunching, accounting for entanglement between $z_I$ and $r_i$ is necessary for quantitative correctness.
\subsubsection{Mobile ion: impact on the ion}\label{sssec:s0_mobile_ion}
\begin{figure}
    \centering
    \includegraphics[width=0.75\columnwidth]{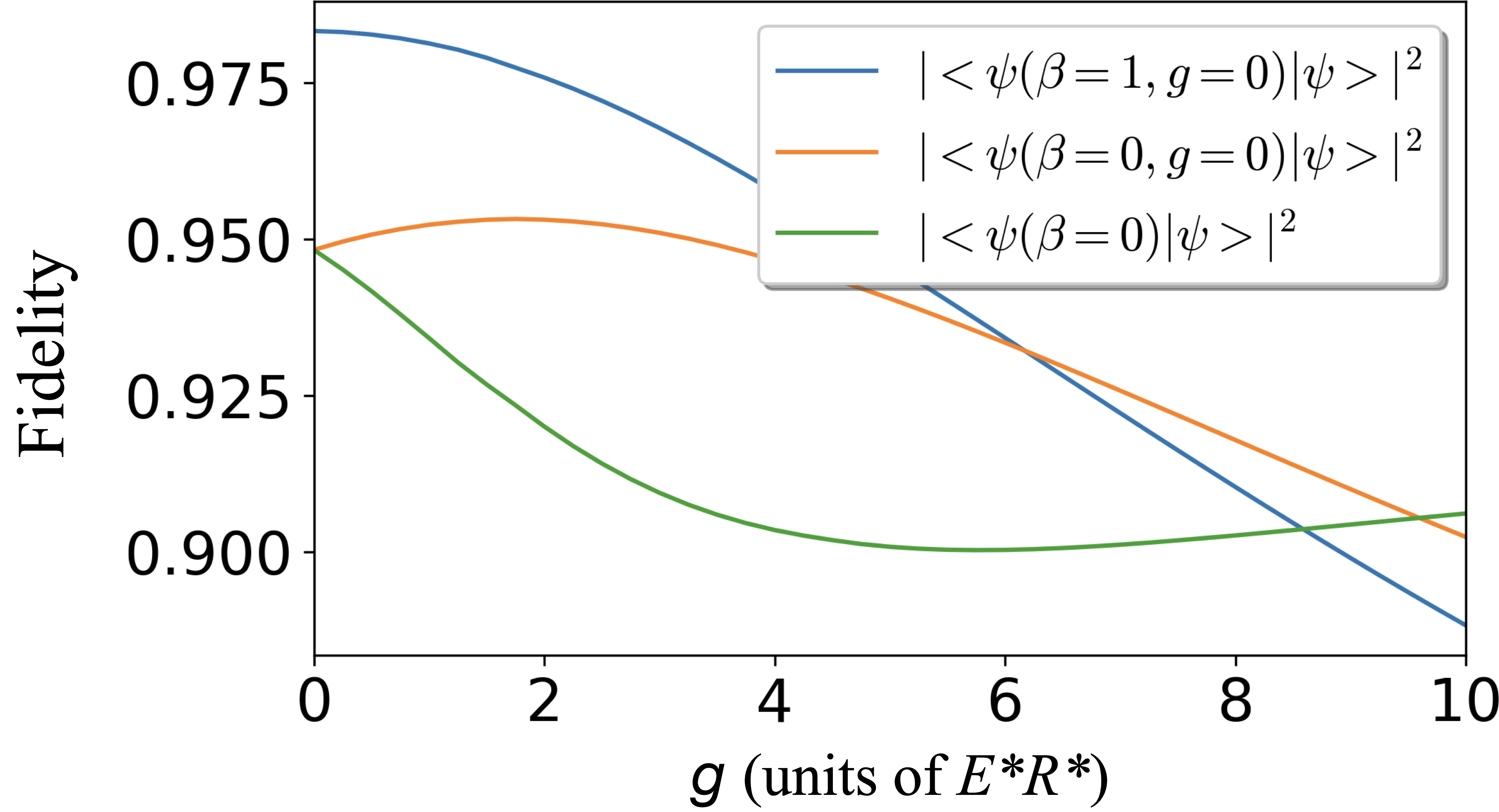}
    \caption{Uhlmann fidelity between the exact ground state $\ket{\psi}$ at $\beta=1$ and the state describing a mobile ion with non-interacting atoms $\ket{\psi(\beta=1,g=0)}$ (blue line), a static ion with non-interacting atoms $\ket{\psi(\beta=0,g=0)}$ (orange line), and a static ion with interacting atoms $\ket{\psi(\beta=0)}$ (green line).}
    \label{fig:fid_s0}
\end{figure}
\begin{figure}
    \centering
    \includegraphics[width=0.75\columnwidth]{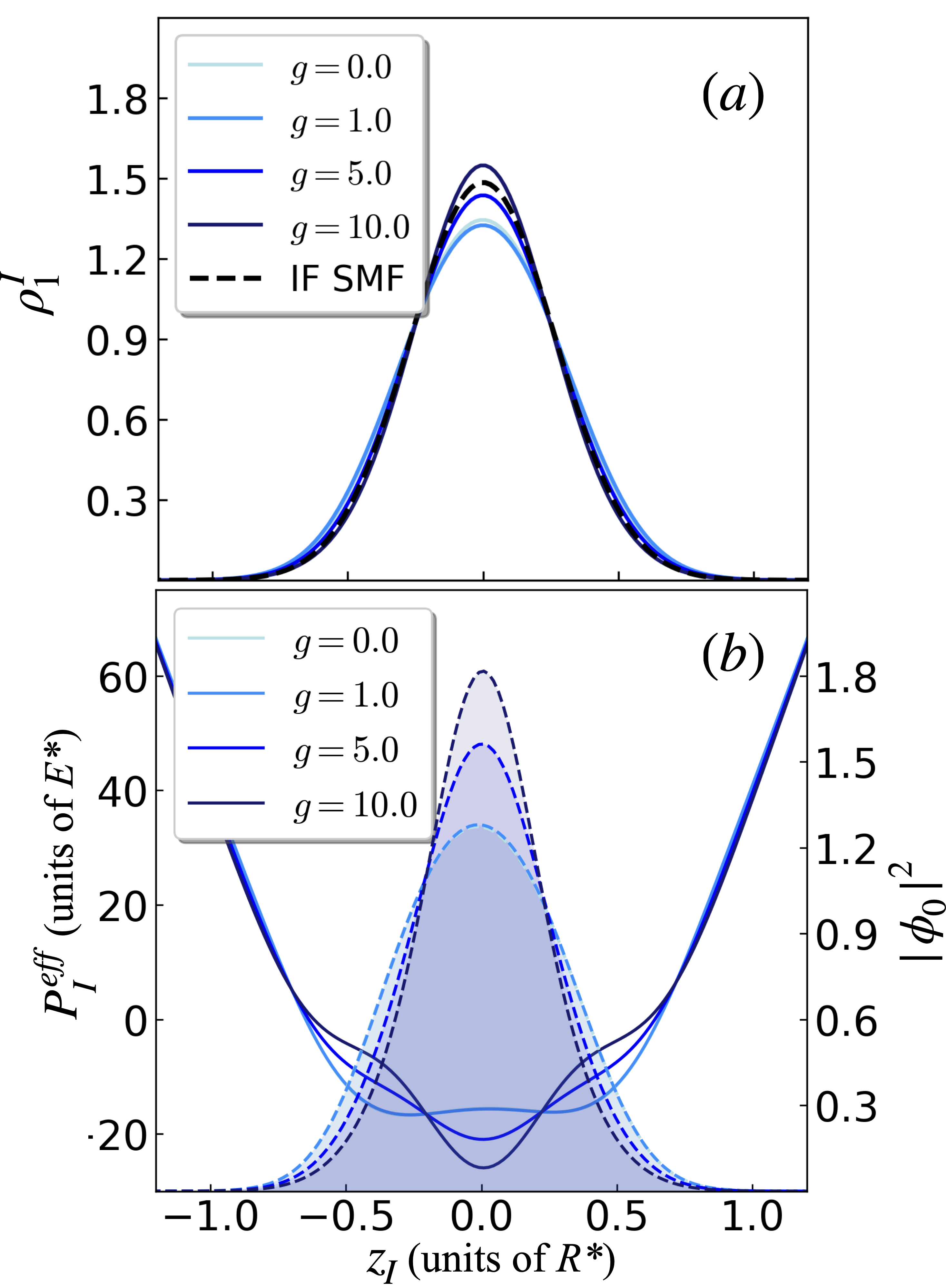}
    \phantomsubfloat{\label{fig:pI}}
    \phantomsubfloat{\label{fig:eff_ion_s0}}
    \vspace{-2\baselineskip}
    \caption{
    (a) the ion density $\rho_1(z_{I})$ at various coupling strengths $g$, obtained via: the ion frame species mean field (IF SMF) ansatz (dashed black line) and full ML-MCTDH ansatz (solid lines). Note the IF SMF has only a single solution for all $g$, since the $z_{I}$ and $r_i$ degrees of freedom decouple in the SMF ansatz (see eq.~\eqref{eq:product_ansatz}). (b) The effective potential $P_I^{eff}(z_{I})$ experienced by the mobile ion due to the exact atomic density $\rho^{IF}_1(z_{A})$ at various coupling strengths $g$. The filled curves denote the corresponding ground state orbitals $|\phi_0|^2$.
    }
    \label{fig:s0_ion_entanglement}
\end{figure}
We now perform a complementary analysis on how the mobile ion 
is affected by the interatomic coupling strength $g$.
In~\cref{fig:ecpt_s0_3},
we observe that the ion's potential energy $P_{I}$ slightly decreases with $g$ (red solid line), 
whilst its kinetic energy $K_{I}$ slightly increases
(blue solid line). 
From this, we can infer that with increasing $g$ the ion density $\rho_1(z_{I})$ 
becomes squeezed and more localised at the trap centre.\\
This observation is further confirmed by examining the Uhlmann fidelity $|\braket{\chi|\psi}|^2$ between the numerically-exact ground state $\ket{\psi}$ for the mobile ion system and several states $\ket{\chi}$ describing different limiting cases (see~\cref{fig:fid_s0}). At weak couplings, $\ket{\psi}$ bears strongest similarity to the
state describing a mobile ion with non-interacting atoms $\ket{\psi(\beta=1,g=0)}$.
Then around $g\approx6$, the dominant overlap is with the state 
describing a static ion with non-interacting atoms $\ket{\psi(\beta=0,g=0)}$,
having no correlations at all.
Finally, for $g>9$ it shares the greatest overlap with the state $\ket{\psi(\beta=0)}$, 
describing a static ion with interacting atoms.
The latter exhibits only atom-atom correlations. \\
To develop a intuitive picture of what is happening to the ion, we first perform a comparison in terms of the SMF ansatz and exact solution in the IF
for the ion density $\rho_1(z_{I})$, similar to what was done for the atoms in~\ref{sssec:s0_mobile_atom}. Note that the SMF prediction for $\rho_1(z_{I})$ is independent of $g$, since here the DOF are decoupled (dashed black line in~\cref{fig:pI}). Thus, the ion localisation phenomenon cannot be captured by the SMF ansatz alone. For weak coupling $g\leq1$, the ion density $\rho_1(z_{I})$ obtained via the exact result spreads out subtly (note the dip in probability at $z_{I}=0$ between $g=0$ and $g=1$), before gradually becoming higher and narrower (see solid curves in~\cref{fig:pI}), in line with our prior observations of $K_{I}$ and $P_{I}$ discussed in the paragraph above. \\
Next, we construct an effective Hamiltonian for the ion due to the atomic density, 
in the same manner as we did for the atoms:
\begin{equation}
    \begin{split}
        H_{I}^{eff} &= K_I + P_I + N\int dz_{A} \; 
        V_{\text{AI}}(z_{A},z_{I}) \rho^{IF}_1(z_{A})\\
        &= K_I + P_I^{eff},
    \end{split}
\end{equation}
where $\rho^{IF}_1(z_{A})$ is the exact one-body density of the atoms 
obtained in the IF.
The effective potential $P_I^{eff}$ is given 
in~\cref{fig:eff_ion_s0} for various interatomic couplings.
The exact one-body densities $\rho_1(z_I)$ fit well inside $P_I^{eff}$. 
At $g=0$, the effective trap takes the form of a harmonic trap 
with a shallow double-well modulation near the origin. 
As the intra-species correlations build up, the double-well structure inverts, 
making the ion profile narrower due to the new minimum at the origin.
\subsection{First excited state}\label{ssec:s1}
In the first excited state, both of the atom-ion bound states are occupied by a single atom. The opposite symmetries of the bound states force the atoms to reside on the same side w.r.t. the ion, such that the atoms share a large spatial-overlap (see~\cref{ssec:s1_static}). The spatial overlap leads to a swift rise in the total energy of the first excited at finite interatomic interactions. To minimise their overlap at greater interaction strengths, one of the atoms is released from the ion and occupies a vibrational trap state. Whilst the ion’s mobility leads to an overall positive shift in the total energy of the state, as well as increased inter- and intra-species separations, it does not qualitatively affect the underlying physics (see~\cref{ssec:s1_mobile}).
\begin{figure*}
    \centering
    \includegraphics[width=1.9\columnwidth]{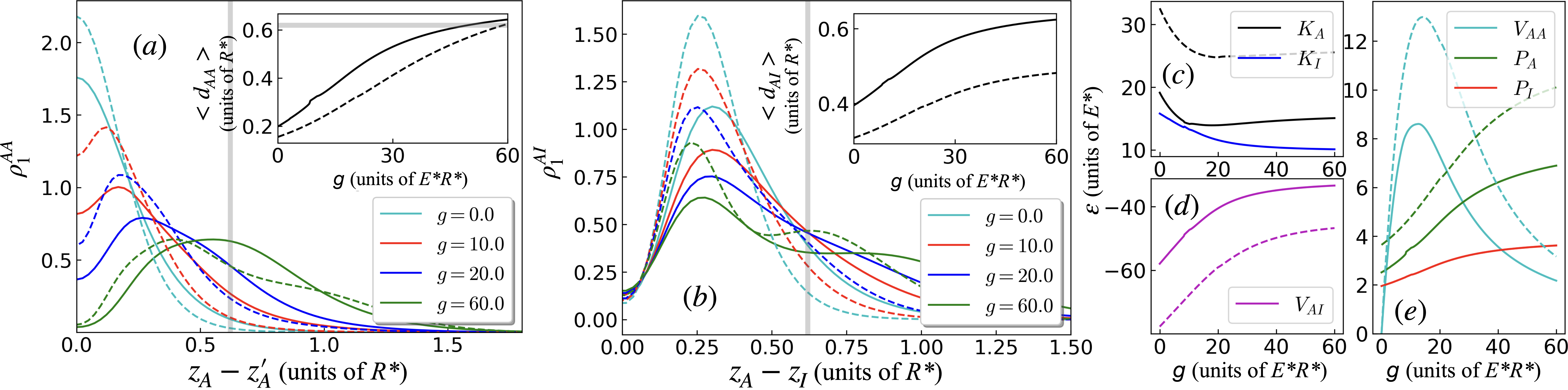}
    \phantomsubfloat{\label{fig:sep_s1}}
    \phantomsubfloat{\label{fig:AI_sep_s1}}
    \phantomsubfloat{\label{fig:ecpt_s1}}
    \phantomsubfloat{\label{fig:ecpt_s1_2}}
    \phantomsubfloat{\label{fig:ecpt_s1_3}}
    \vspace{-2\baselineskip}
    \caption{Key observables for the first excited state. \textbf{(a):} interatomic separation distribution $\rho^{AA}_1(z_{A} - z_{A}^{\prime})$ for different atom-atom coupling strengths $g$. The inset shows the expectation value $\braket{d_{AA}}$ for the atom-atom separation as a function of $g$ (eq.~\eqref{eq:d_AA}).  \textbf{(b):} interspecies separation distribution $\rho^{AI}_1(z_{A} - z_{I})$ for varying atom-atom coupling strengths $g$. The inset shows the expectation values $\braket{d_{AI}}$ for the atom-ion separation as a function of $g$ (eq.~\eqref{eq:d_AI}). \textbf{(c)-(e):} the evolution of the laboratory frame energy components with atom-atom coupling strength $g$. \textit{Note for (a) and (b):} the grey lines indicate the distance between the minima of the atom-ion interaction potential \eqref{eq:atom-ion-int}. Due to the parity symmetry it is sufficient to show only the positive semi-axis. \textit{All subfigures:} the solid curves correspond to a mobile ion, whilst dashed curves correspond to a static ion.}
    \label{fig:s1_observables}
\end{figure*}
\begin{figure*}
    \centering
    \includegraphics[width=1.35\columnwidth]{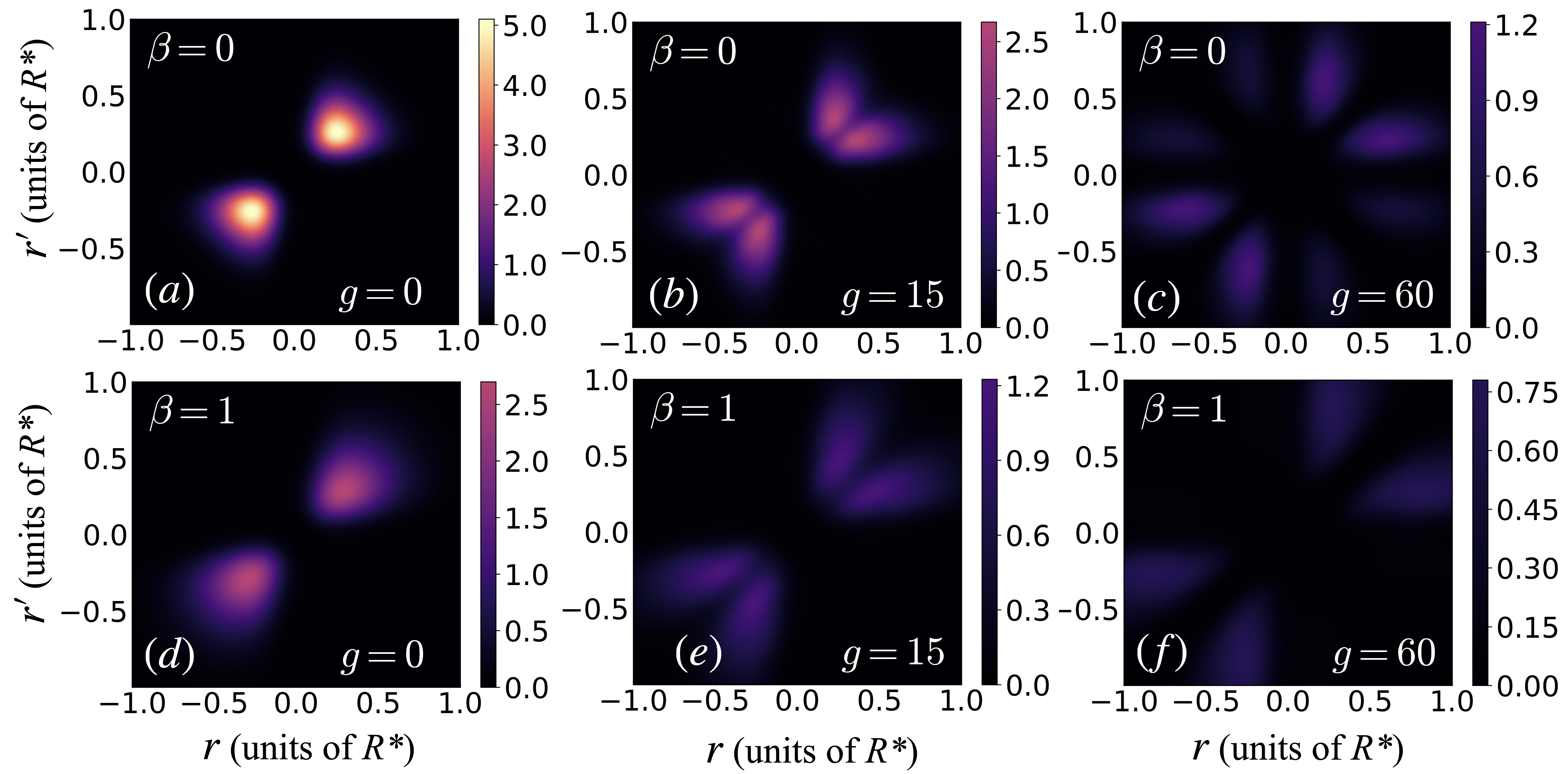}
    \phantomsubfloat{\label{fig:dmat2_s1_1}}
    \phantomsubfloat{\label{fig:dmat2_s1_2}}
    \phantomsubfloat{\label{fig:dmat2_s1_3}}
    \phantomsubfloat{\label{fig:dmat2_s1_4}}
    \phantomsubfloat{\label{fig:dmat2_s1_5}}
    \phantomsubfloat{\label{fig:dmat2_s1_6}}
    \vspace{-2\baselineskip}
    \caption{
        Snapshots of the atomic probability density $\rho_2^{\text{AA}} = |\psi(r,r^{\prime})|^2$ of the first excited state for different interaction strengths $g$ for a static ion ((a) to (c)) and a mobile ion ((d) to (f)).
    }
    \label{fig:dmat2_s1}
\end{figure*}

\begin{figure}
    \centering
     \includegraphics[width=0.75\columnwidth]{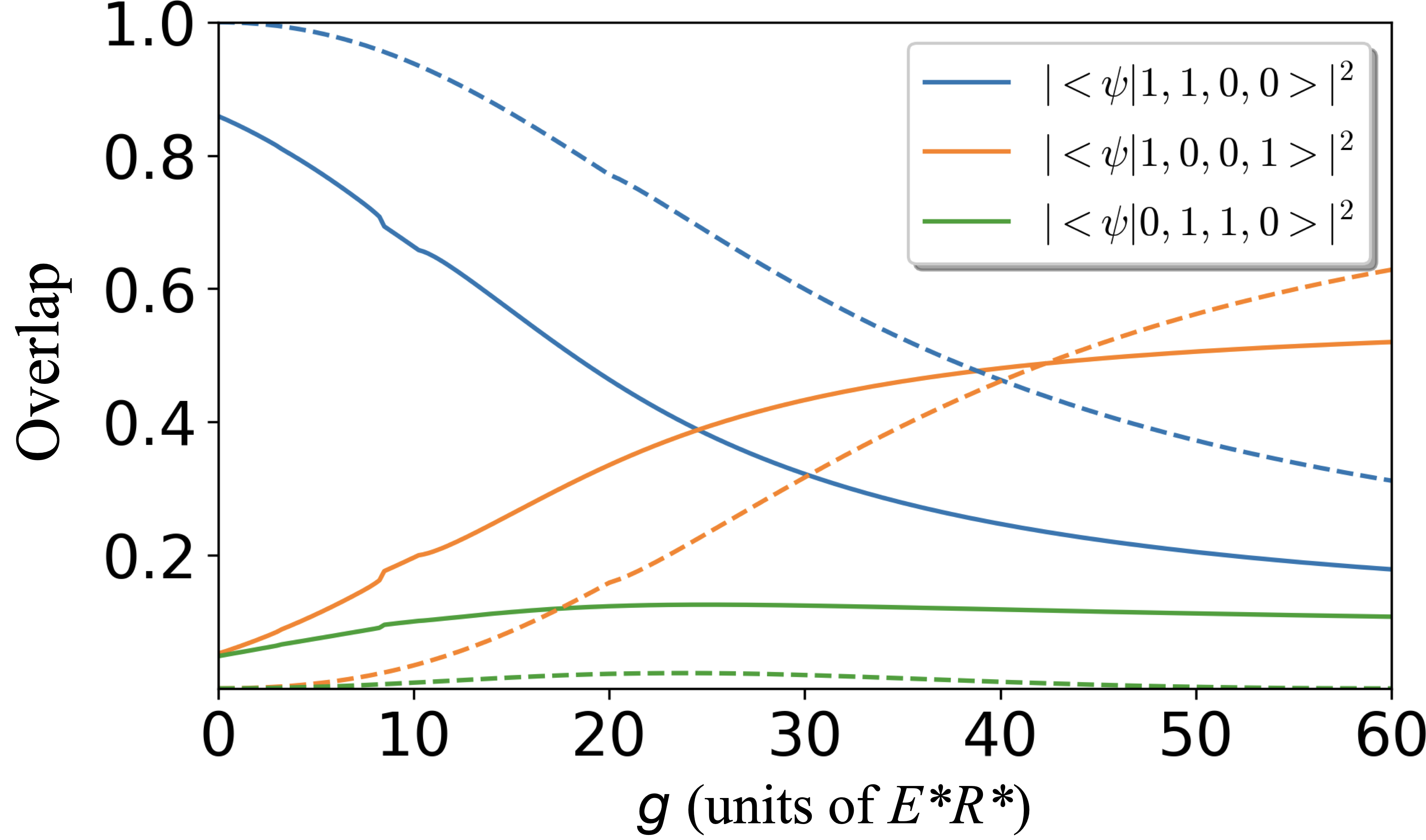}
     \caption{
     Overlap spectrum $|\braket{\psi|\textbf{n}}|^2$ between the first excited state $\ket{\psi}$ and the separate SPF number states $\ket{\textbf{n}}$ calculated from the static ion model in eq.~\eqref{eq:h1b} for a static ion (dashed curves) and a mobile ion (solid curves). For the sake of clarity, we show here only the dominant overlap coefficients. The complete representation of the mobile ion state in the static ion basis requires numerous additional small contributions from higher-order number states.
     }
     \label{fig:s1_ns_spec}
\end{figure}
\subsubsection{Static ion}\label{ssec:s1_static}
The first excited state for two non-interacting ($g=0$) atoms 
coupled to a static ($\beta=0$) ion is given by the number state $\ket{1,1,0,0}$ w.r.t.\ the single-particle eigenstates of $h_{\text{1b}}$ (see eq.~\eqref{eq:h1b}), 
i.e. each molecular orbital $\phi_0,\; \phi_1$ in~\cref{fig:h1b_solutions} is occupied by a single atom. 
Unlike in the ground state, the atoms here are completely bunched (note the single peak in~\cref{fig:sep_s1}). 
This is caused by the fact that $\phi_0(r) \approx -\sign(r) \phi_1(r)$ (compare $\phi_0$ and $\phi_1$ in~\cref{fig:h1b_solutions}). 
The suppression of the anti-bunched probabilities can be seen explicitly by inserting this relation into the two-body density:
\begin{equation}\label{eq:s1_density}
\begin{split}
    \rho_2(r,r') =  &\frac{1}{4}\big(|\phi_0(r)|^2|\phi_1(r')|^2 + |\phi_1(r)|^2|\phi_0(r')|^2 \\
    &+ 2\phi_0^*(r)\phi_1(r)\phi_1^*(r')\phi_0(r')\big) + \text{c.c.},
\end{split}
\end{equation}
where the first two terms cancel out the last two terms whenever $\sign(r) \neq \sign(r')$.\\
With increasing $g$, the interatomic separation distribution $\rho_1^{AA}$ spreads 
and the peak at $z_A=z'_A$ for $g=0$ is shifted to larger distances 
(dashed curves in~\cref{fig:sep_s1}).
Even though the average distance $d_{AA}$ between the atoms 
gradually increases (black dashed curve in the inset of~\cref{fig:sep_s1}), 
they insist on staying in the bunched configuration, 
i.e. on the same side w.r.t.\ ion, which results in the formation of a nodal structure on the diagonal of the probability density (see~\cref{fig:dmat2_s1_2,fig:dmat2_s1_3}). Thus, the inter-species separation distribution peak broadens (dashed curves in~\cref{fig:AI_sep_s1}) which leads to a monotonous increase of $d_{AI}$ 
(black dashed curve in the inset of~\cref{fig:AI_sep_s1}).\\
The above observations clarify the rapid increase of the total energy with $g$
(see~\cref{fig:static_spec_ext}). 
One major contribution comes from the atom-atom interaction energy $V_{AA}$
given by $g \rho_1^{AA}(r=0)$. 
Thus, considering a large initial amplitude $\rho_1^{AA}(r=0)>1$ at $g=0$
and that it decreases slowly with $g$, 
we identify a fast linear increase of $V_{AA}$ up to approximately $g=20$ 
(light blue dashed line in~\cref{fig:ecpt_s1_3}).
Once the amplitude drops significantly below the value of $1.0$,
$V_{AA}$ starts decreasing.
Another significant contribution stems from the atom-ion interaction potential $V_{AI}$.
Drawing away from the ion requires the atoms
to climb the $V_{AI}$ potential,
which costs energy (purple dashed line in ~\cref{fig:ecpt_s1_2}).
For $g<10$, this energy is compensated 
by decreasing atomic kinetic energy $K_A$
(black dashed line in ~\cref{fig:ecpt_s1}), 
but $V_{AI}$ keeps increasing even after $K_A$ has saturated.\\
That the atoms stay on one side of the ion 
while increasing their separation with $g$
is due to the growing contribution of the number state $\ket{1,0,0,1}$ 
(see dashed lines in~\cref{fig:s1_ns_spec}).  
The eigenstate is continuously transitioning to a regime 
in which one atom remains bound to the ion
and the other is released into the harmonic trap. This can be seen clearly from the additional peak around $\approx\pm0.7R^*$ that emerges in the interspecies separation distribution (see green dashed curve in~\cref{fig:AI_sep_s1}).
This causes a monotonous increase of the atomic trap potential energy $P_A$ 
(green dashed line in~\cref{fig:ecpt_s1_3}).
Beyond $g=40$, the overlap with $\ket{1,0,0,1}$ becomes dominant.
Since the odd vibration orbital $\phi_3$ features a smaller probability density 
to be found at the minimum of the atom-ion potential than $\phi_0$,
we observe an emergence of anti-bunching probability in the upper-left and lower-right quadrants of $\rho_2$
(see~\cref{fig:dmat2_s1_3}). 
\subsubsection{Mobile ion}\label{ssec:s1_mobile}
The ion's mobility causes a by-mixture of the number state $\ket{0,1,1,0}$
(green solid line in~\cref{fig:s1_ns_spec}), 
whose contribution to the eigenstate amounts to $\sim 10\%$, 
and is approximately unchanged by the atom-atom coupling strength $g$.
Thus, the impact on the physical quantities from the static ion case
is expected to be qualitatively similar at all $g$.\\
The positive off-set of the total energy of the first excited state (see~\cref{fig:spectrum})
is mainly due to the energy of the ion itself $K_I+P_I$
(dark blue and red solid lines in~\cref{fig:ecpt_s1,fig:ecpt_s1_3}).
With increasing $g$, there is an exchange of energy between $K_I$ and $P_I$,
with $K_{I}$ decreasing and $P_I$ increasing,
indicating the delocalisation of the ion.
The ion's mobility induces an additional energy exchange between $K_A$ and $V_{AI}$,
with $V_{AI}$ increasing and $K_A$ decreasing
(compare black and purple solid curves in~\cref{fig:ecpt_s1,fig:ecpt_s1_2} to dashed ones),
implying that the atoms separate from each other.
This is also evident from the patterns of $\rho_2(r,r')$, 
which remain qualitatively the same, albeit
with a slightly enhanced overall spread
(compare rows in~\cref{fig:dmat2_s1}) that is further imprinted on the interatomic $\rho_1^{AA}$ and interspecies $\rho_1^{AI}$ separation distributions 
(compare dashed and solid lines in~\cref{fig:sep_s1,fig:AI_sep_s1}).
Accordingly, the distance among the atoms $d_{AA}$ increases 
(black lines in the inset of~\cref{fig:sep_s1}).
This is in contrast to the ground state, 
where the ion's mobility decreased $d_{AA}$.
Due to the overall decreased amplitude of $\rho_1^{AA}(r=0)$ compared to $\beta=0$,
the atom-atom interaction energy $V_{AA}$ reaches the turning point already 
at a slightly weaker coupling $g$ 
(light blue solid line in~\cref{fig:ecpt_s1_3}).
In summary, at $\beta=1$ the atoms separate
further from each other and from the ion as compared to $\beta=0$.
\subsection{Second excited state}\label{ssec:s2}
In the non-interacting second excited state, both atoms occupy the upper bound state in the atom-ion interaction potential and display preference neither for bunching nor anti-bunching (see~\cref{ssec:s2_static}). The response of the second excited state to finite atomic interactions closely resembles that of the first excited state, namely that at intermediate interactions, the atoms become preferentially bunched and at large interaction strengths, one of the atoms frees itself from the ion. In addition, the ion’s mobility does not produce significant qualitative differences to the results for the static case (see~\cref{ssec:s2_mobile}). These similarities between the states are to be expected, considering that their energy gap initially narrows (see~\cref{fig:static_spec_ext}). There are however, two crucial differences between the first and second excited states: (i) the bunching of the atoms in the latter state does not stem from an inherent symmetry of the single particle states, but rather is a consequence of state mixing with other number states, such as $\ket{2,0,0,0}$ and (ii) the atomic kinetic energy is consistently higher in the second excited state.
\begin{figure*}
    \centering
    \includegraphics[width=1.9\columnwidth]{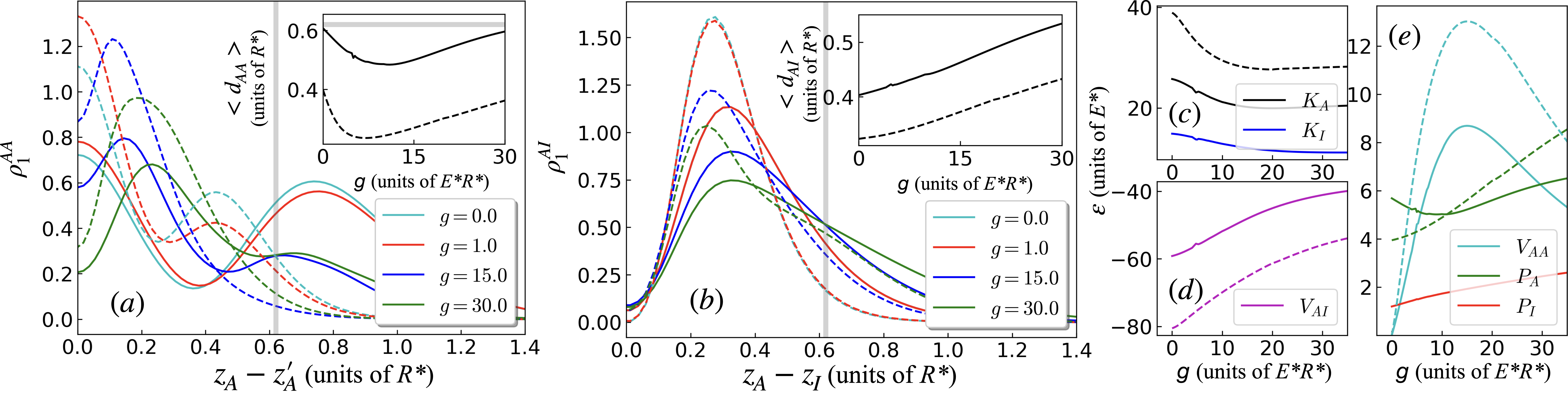}
    \phantomsubfloat{\label{fig:sep_s2}}
    \phantomsubfloat{\label{fig:AI_sep_s2}}
    \phantomsubfloat{\label{fig:ecpt_s2}}
    \phantomsubfloat{\label{fig:ecpt_s2_2}}
    \phantomsubfloat{\label{fig:ecpt_s2_3}}
    \vspace{-2\baselineskip}
    \caption{Key observables for the second excited state. \textbf{(a):} interatomic separation distribution $\rho^{AA}_1(z_{A} - z_{A}^{\prime})$ for different atom-atom coupling strengths $g$. The inset shows the expectation value $\braket{d_{AA}}$ for the atom-atom separation as a function of $g$ (eq.~\eqref{eq:d_AA}).  \textbf{(b):} interspecies separation distribution $\rho^{AI}_1(z_{A} - z_{I})$ for varying atom-atom coupling strengths $g$. The inset shows the expectation values $\braket{d_{AI}}$ for the atom-ion separation as a function of $g$ (eq.~\eqref{eq:d_AI}). \textbf{(c)-(e):} the evolution of the laboratory frame energy components with atom-atom coupling strength $g$. \textit{Note for (a) and (b):} the grey lines indicate the distance between the minima of the atom-ion interaction potential \eqref{eq:atom-ion-int}. Due to the parity symmetry it is sufficient to show only the positive semi-axis. \textit{All subfigures:} the solid curves correspond to a mobile ion, whilst dashed curves correspond to a static ion.
    }
    \label{fig:s2_observables}   
\end{figure*}
\begin{figure*}
    \centering
    \includegraphics[width=1.35\columnwidth]{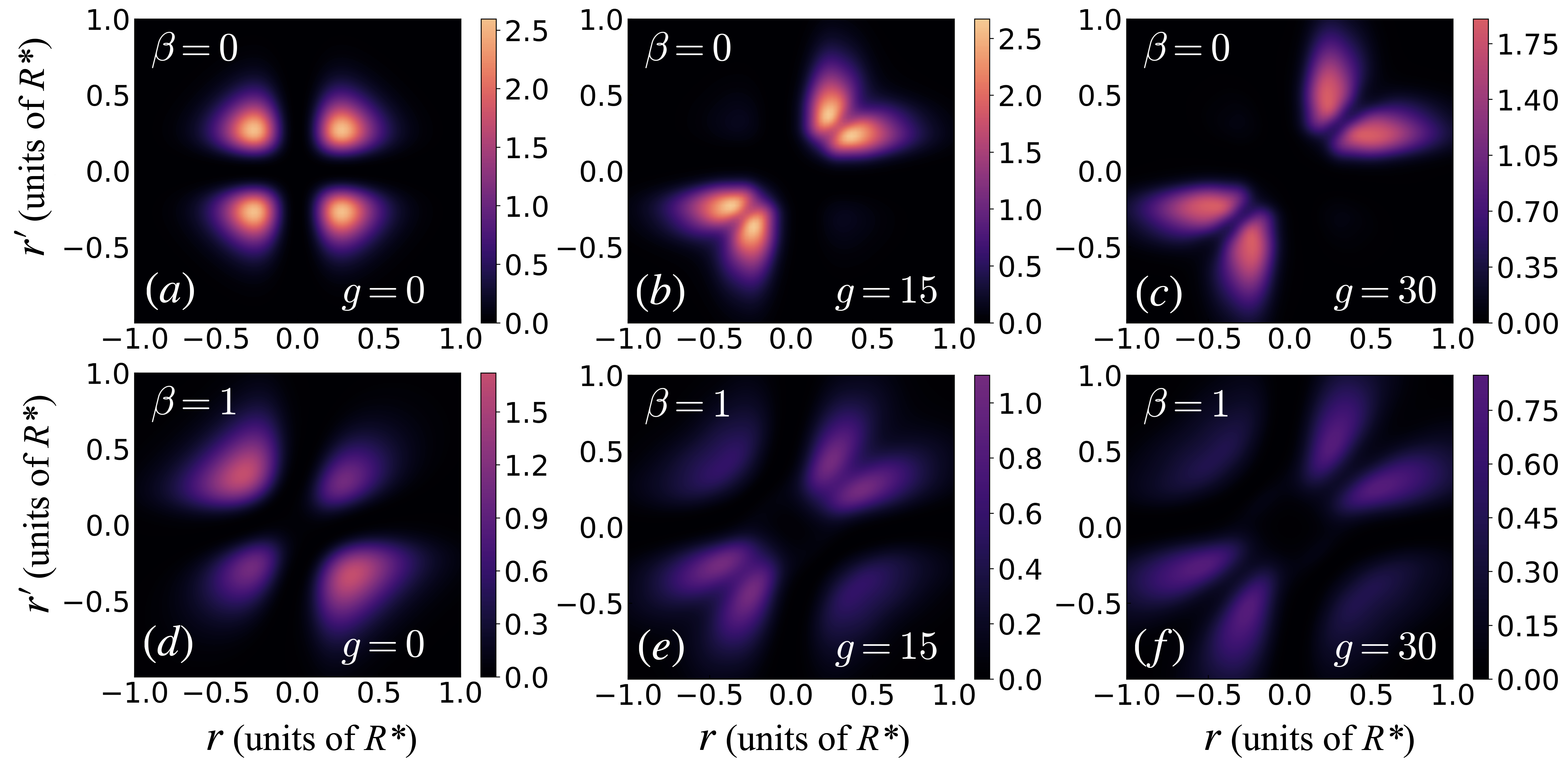}
    \phantomsubfloat{\label{fig:dmat2_s2_1}}
    \phantomsubfloat{\label{fig:dmat2_s2_2}}
    \phantomsubfloat{\label{fig:dmat2_s2_3}}
    \phantomsubfloat{\label{fig:dmat2_s2_4}}
    \phantomsubfloat{\label{fig:dmat2_s2_5}}
    \phantomsubfloat{\label{fig:dmat2_s2_6}}
    \vspace{-2\baselineskip}
    \caption{
        Snapshots of the atomic probability density $\rho_2^{\text{AA}} = |\psi(r,r^{\prime})|^2$ of the second excited state for different interaction strengths $g$ for a static ion ((a) to (c)) and a mobile ion ((d) to (f)).
    }
    \label{fig:dmat2_s2}
\end{figure*}

\begin{figure}
    \centering
     \includegraphics[width=0.75\columnwidth]{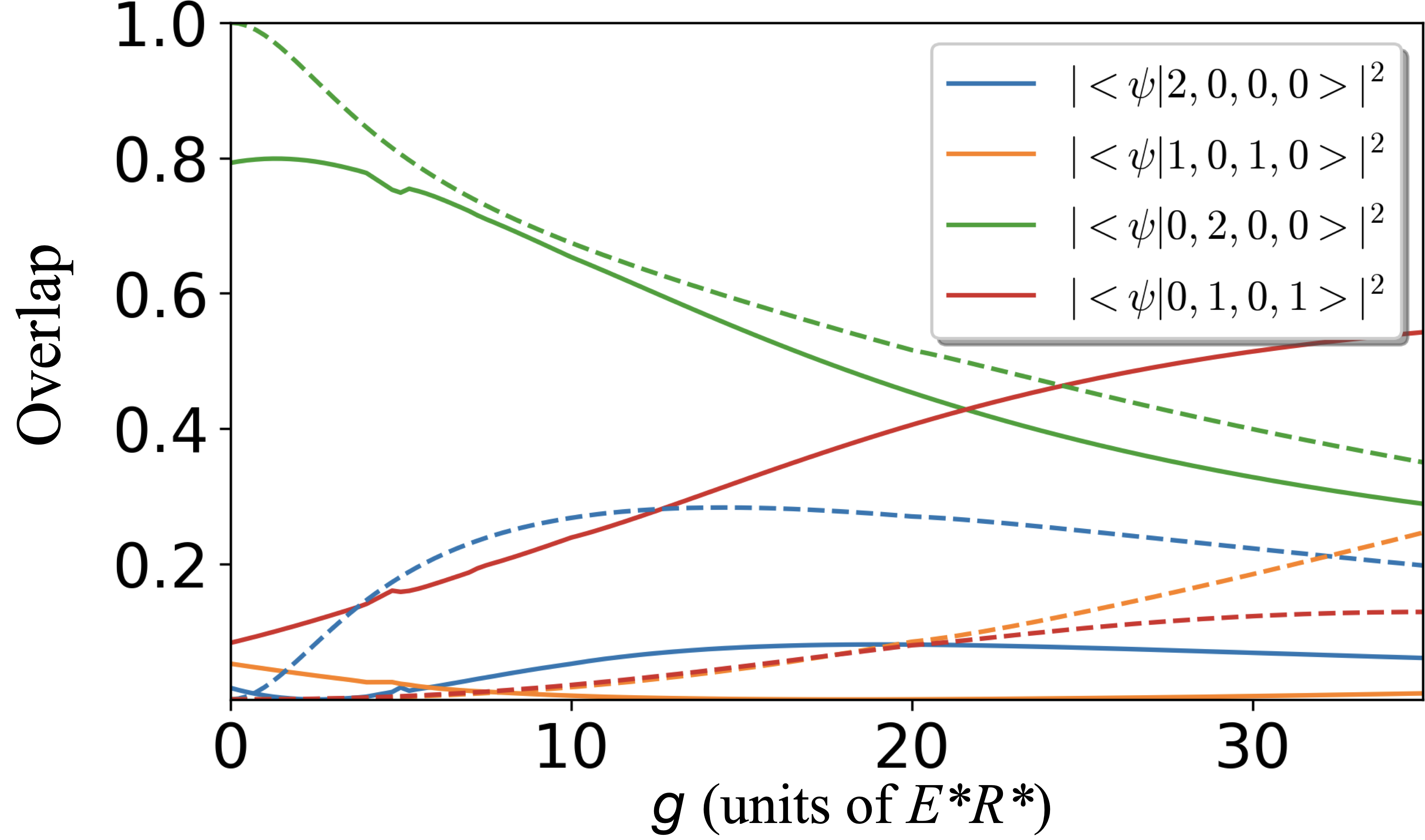}
     \caption{
     Overlap spectrum $|\braket{\psi|\textbf{n}}|^2$ between the second excited state $\ket{\psi}$ and the separate SPF number states $\ket{\textbf{n}}$ calculated from the static ion model in eq.~\eqref{eq:h1b} for a static ion (dashed curves) and a mobile ion (solid curves). For the sake of clarity, we show here only the dominant overlap coefficients. The complete representation of the mobile ion state in the static ion basis requires numerous additional small contributions from higher-order number states.
     }
     \label{fig:s2_ns_spec}
\end{figure}
\subsubsection{Static ion}\label{ssec:s2_static}
The second excited state for two non-interacting ($g=0$) atoms  
coupled to a static ($\beta=0$) ion is given by the number state $\ket{0,2,0,0}$ 
w.r.t.\ the single-particle eigenstates of $h_{\text{1b}}$ (see eq.~\eqref{eq:h1b}), 
with both atoms occupying the molecular orbital $\phi_1$ 
(see~\cref{fig:h1b_solutions}). 
Similarly to the ground state, 
the atoms show no preference for either bunched 
or anti-bunched configurations (see~\cref{fig:dmat2_s2_1}).\\
As one increases $g$, there is 
a by-mixture of the number state $\ket{2,0,0,0}$ up to $g=15$
(blue dashed line in~\cref{fig:s2_ns_spec}). 
This leads to an increased bunching of the atoms 
(red dashed line in~\cref{fig:sep_s2}),
decreasing the average distance $d_{AA}$ among them
(black dashed line in the inset of~\cref{fig:sep_s2})
until finally the anti-bunched configuration 
is completely suppressed and a nodal structure emerges on the diagonal of the probability density (see~\cref{fig:dmat2_s2_2}). The increasingly-bunched repulsive atoms create an increase in the interspecies separation $d_{AI}$ (see dashed black line in inset of~\cref{fig:AI_sep_s2}).
Beyond $g>10$, we observe a by-mixture 
of the number states $\ket{1,0,1,0}$ and $\ket{0,1,0,1}$ 
(orange and red dashed lines in~\cref{fig:s2_ns_spec})
featuring stronger separations among the atoms,
which causes them to draw away from each other (see the observed increase of $\braket{d_{AA}}$ in the inset of~\cref{fig:sep_s2}). The number states $\ket{1,0,1,0}$ and $\ket{0,1,0,1}$ feature one of the atoms unbound from the ion in a trap state. As such, the interspecies separation distribution broadens significantly (see dashed curves in~\cref{fig:AI_sep_s1}).
At the same time, the contribution of $\ket{2,0,0,0}$ 
displays only a slight decay.
Consequently at larger $g$, we anticipate the preservation of the bunched configuration with an increased spread of the two-body density $\rho_2(r,r')$
and a depletion of the diagonal at $r=r'$.
(see~\cref{fig:dmat2_s2_3}).\\
We note a strong similarity between the atomic probability densities shown in~\cref{fig:dmat2_s2_2,fig:dmat2_s1_2}, 
which is due to a small energy gap between the first and second excited states
(see~\cref{fig:spectrum}).
The two eigenstates also display a similar energy dependence
on the coupling $g$ (compare~\cref{fig:ecpt_s1,fig:ecpt_s1_2,fig:ecpt_s1_3,fig:ecpt_s2,fig:ecpt_s2_2,fig:ecpt_s2_3}). 
The major difference between them is that
the second excited state
has a greater atomic kinetic energy $K_A$.
\subsubsection{Mobile ion}~\label{ssec:s2_mobile}
Similarly to the ground state and first excited state, the ion's mobility leads to 
a shift in the energies 
(compare solid and dashed curves in~\cref{fig:spectrum,fig:ecpt_s2,fig:ecpt_s2_2,fig:ecpt_s2_3})
and of the interatomic and interspecies separations (inset of~\cref{fig:sep_s2,fig:AI_sep_s2}).
Specifically, the atoms exhibit an increase in separation between each other 
of $\sim 0.2 R^*$ and an increase in separation to the ion by $\sim 0.1 R^*$.
Contrary to the first excited state however, 
which featured a $g$-independent by-mixture of an additional number state,
the number state composition of the second excited state 
undergoes substantial structural changes at $\beta=1$ (see~\cref{fig:s2_ns_spec}).
Thus, the role of $\ket{1,0,1,0}$ and $\ket{2,0,0,0}$ 
is substantially suppressed in favour of $\ket{0,1,0,1}$.
As a result, we observe a strong enhancement of the anti-bunched off-diagonal probability in 
$\rho_2(r,r')$ at all $g$ (compare rows in~\cref{fig:dmat2_s2}). 
\subsection{Third excited state}\label{ssec:s3}
In the third excited state, one atom remains bound to the ion in the lowest-energy bound state and the other occupies the lowest-energy trap state. As shown in~\cref{sec:results_spectrum}, the total energy of this eigenstate is not markedly affected by varying atomic interactions. Unsurprisingly, the principle observables for the state (probability density, energy components and number state composition) are likewise largely unaffected by these changes (see~\cref{ssec:s3_static}). This robustness of the third excited state stems from the relatively small spatial-overlap of the two atoms, due to the contrasting length scales of the bound and trap states (see discussion in~\cref{ssec:model}). The state is further robust to the ion mobility (see~\cref{ssec:s3_mobile}), displaying only a positive shift in the total energy due to an increased spread between the atomic and ionic species, in accordance with the weaker localisation of the ion.
\begin{figure*}
    \centering
    \includegraphics[width=1.9\columnwidth]{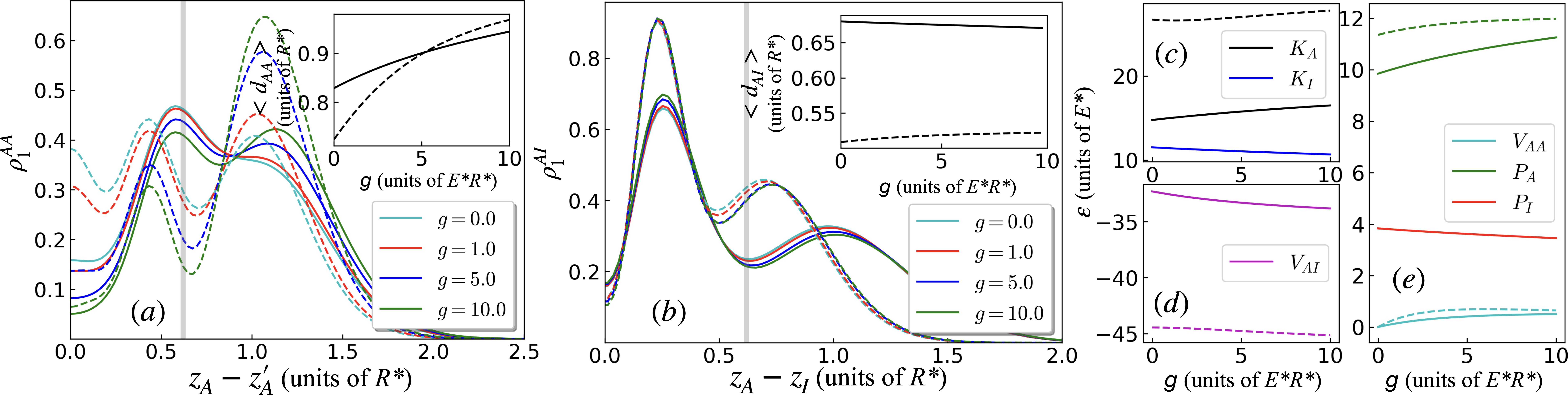}
    \phantomsubfloat{\label{fig:sep_s3}}
    \phantomsubfloat{\label{fig:AI_sep_s3}}
    \phantomsubfloat{\label{fig:ecpt_s3}}
    \phantomsubfloat{\label{fig:ecpt_s3_2}}
    \phantomsubfloat{\label{fig:ecpt_s3_3}}
    \vspace{-2\baselineskip}
    \caption{Key observables for the third excited state. \textbf{(a):} interatomic separation distribution $\rho^{AA}_1(z_{A} - z_{A}^{\prime})$ for different atom-atom coupling strengths $g$. The inset shows the expectation value $\braket{d_{AA}}$ for the atom-atom separation as a function of $g$ (eq.~\eqref{eq:d_AA}).  \textbf{(b):} interspecies separation distribution $\rho^{AI}_1(z_{A} - z_{I})$ for varying atom-atom coupling strengths $g$. The inset shows the expectation values $\braket{d_{AI}}$ for the atom-ion separation as a function of $g$ (eq.~\eqref{eq:d_AI}). \textbf{(c)-(e):} the evolution of the laboratory frame energy components with atom-atom coupling strength $g$. \textit{Note for (a) and (b):} the grey lines indicate the distance between the minima of the atom-ion interaction potential \eqref{eq:atom-ion-int}. Due to the parity symmetry it is sufficient to show only the positive semi-axis. \textit{All subfigures:} the solid curves correspond to a mobile ion, whilst dashed curves correspond to a static ion.
    }
     \label{fig:s3_observables}   
\end{figure*}
\begin{figure*}
    \centering
    \includegraphics[width=1.8\columnwidth]{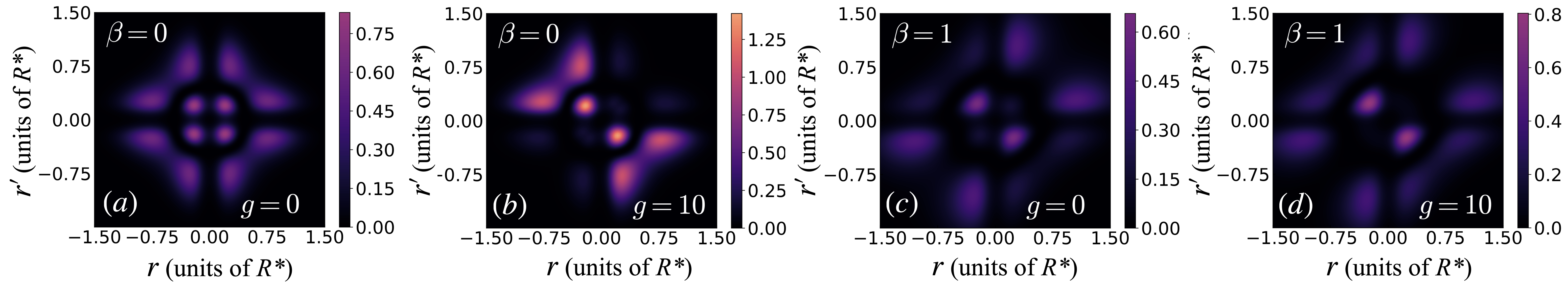}
    \phantomsubfloat{\label{fig:dmat2_s3_1}}
    \phantomsubfloat{\label{fig:dmat2_s3_2}}
    \phantomsubfloat{\label{fig:dmat2_s3_3}}
    \phantomsubfloat{\label{fig:dmat2_s3_4}}
    \vspace{-2\baselineskip}
    \caption{
        Snapshots of the atomic probability density $\rho_2^{\text{AA}} = |\psi(r,r^{\prime})|^2$ of the third excited state for different interaction strengths $g$ for a static ion ((a) and (b)) and a mobile ion ((c) and (d)).
    }
    \label{fig:dmat2_s3}
\end{figure*}

\subsubsection{Static ion}~\label{ssec:s3_static}
The third excited state for two non-interacting ($g=0$) atoms  
coupled to a static ($\beta=0$) ion is given by the number state $\ket{1,0,1,0}$,
where one atom is bound in the even molecular orbital $\phi_0$ 
and the other occupies the even vibrational orbital $\phi_2$ 
(see~\cref{fig:h1b_solutions}).\\
Previously, we have seen that the total energy of the vibrational eigenstates
depends only weakly on $g$ (see~\cref{fig:spectrum}).
This also holds for the energy components, 
which show only a slight exchange between $V_{AI}$ and $K_A$ (see~\cref{fig:ecpt_s3,fig:ecpt_s3_2}).
On the level of energies, the eigenstate seems quite robust 
to perturbations by atom-atom interactions, though the atom-atom separation distribution $\rho_1^{AA}$ features significant structural changes with increasing $g$
leading to increased separation among the atoms $d_{AA}$ 
(black dashed curve in the inset of~\cref{fig:sep_s3}).\\
At $g=0$, there are three pronounced humps in $\rho_1^{AA}$
(light blue dashed line in~\cref{fig:sep_s3}).
Comparing this to $\rho_2(r,r')$ (\cref{fig:dmat2_s3_1}), we can see that these correspond to: (i) atoms 
being bound to the same side of the ion ($r_A=z_A-z'_A=0$), (ii) atoms
being bound to opposite sides of the ion ($r_A \sim 0.4 R^*$), (iii)
and one atom being bound to the ion whilst the other is unbound ($r_A \sim 1.0R^*$).
With increasing $g$, we observe a depletion of the central peak at $r_A=0$
in favour of the outer peak, which additionally shifts to larger separations
(dashed curves in~\cref{fig:sep_s3}). The middle peak is essentially unaffected.
The bunching of atoms becomes suppressed and at $g=10$ one ends up
with atoms located on different sides w.r.t.\ the ion (see \cref{fig:dmat2_s3_2}).\\
The structure of $\rho_1^{AI}$ at $g=0$ has two distinct peaks at $\approx0.3R^*$ and $\approx0.8R^*$, reflecting the different length scales of the molecular and vibrational orbitals comprising the number state (dashed curves in~\cref{fig:AI_sep_s3}). $\rho_1^{AI}$ is largely unaffected by the interatomic coupling strength, though there is a peak shift around $\approx0.8R^*$, indicating that the increase in $d_{AI}$ (dashed curve in inset of~\cref{fig:AI_sep_s3}) arises solely from the unbound atom spreading out as the atom pair grows increasingly repulsive.
\subsubsection{Mobile ion}~\label{ssec:s3_mobile}
Several patterns in $\rho_2(r,r')$ are enhanced for $\beta=1$
(compare rows in~\cref{fig:dmat2_s3}). 
When both atoms are in the vicinity of the ion ($r<0.3$, $r'<0.3$) 
the anti-bunching is amplified, while when one atom is further away,
($r \lessgtr 0.3R^*$, $r' \gtrless 0.3R^*$)
the anti-bunching is suppressed and atoms are most likely to be found
on the same side w.r.t\ the ion.
This causes further depletion of the interparticle separation distribution 
at $r_A=0$
and suppresses the outer peak at $r_A \sim 1.0 R^*$
(compare dashed and solid lines in~\cref{fig:sep_s3}).
The relative magnitudes of these effects are different 
depending on the value of $g$, resulting in larger $d_{AA}$ 
at small $g$ and smaller $d_{AA}$ at large $g$, compared to the static ion case
(black curves in the inset of~\cref{fig:sep_s3}). The ion's mobility creates a positive shift in $d_{AI}$ from the greater spread of $\rho_1^{AI}$. Although the vibrational orbital peak shifts to greater separations, the molecular orbital peak at $\approx0.3R^*$ is enhanced, leading to a slow decrease of $d_{AI}$.
\subsection{Fourth excited state}\label{ssec:s4}
The non-interacting fourth excited state bears similarity to the third excited state, except that the trap-state atom occupies the next highest-energy vibrational orbital. Here, the opposing symmetries of the single particle states suppress the off-diagonal elements ($z_A = -z_A^{\prime}$) of the probability density, as was observed already in the first excited state. For the case of a static ion, the atoms fully-separate to opposite sides of the ion even at extremely weak atomic interaction strengths and thereafter, further changes are negligible (see~\cref{ssec:s4_static}). For the case of an equal mass system however, the ion’s mobility reinforces the overlap of the atoms, which competes against the anti-correlations produced by the repulsive interatomic interaction. As a result, the total energy of the state grows rapidly and the on-set of the fully anti-bunched configuration is delayed (see~\cref{ssec:s4_mobile}).
\begin{figure*}
    \centering
    \includegraphics[width=1.9\columnwidth]{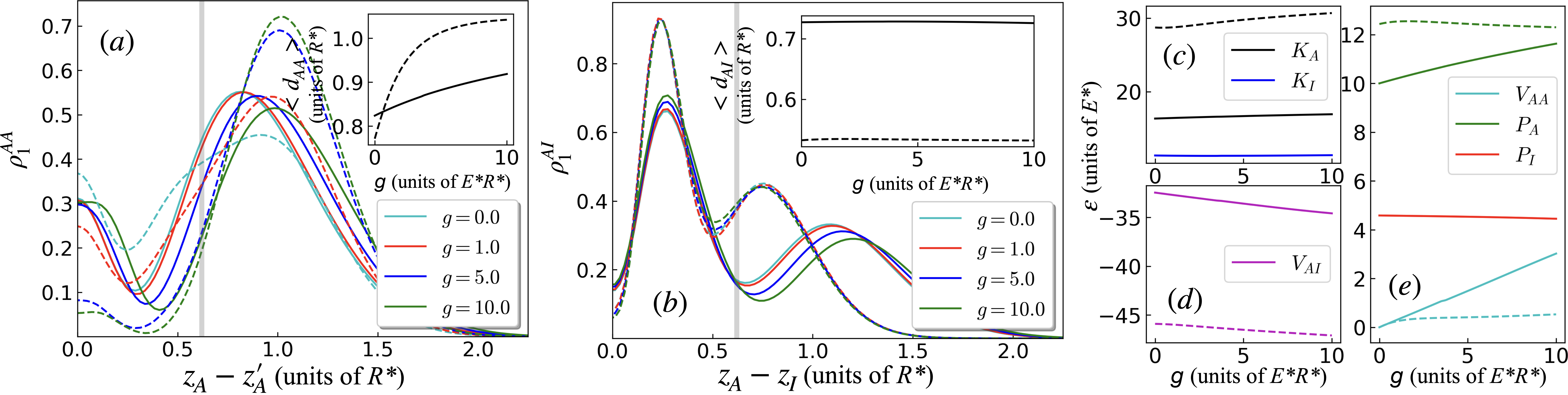}
    \phantomsubfloat{\label{fig:sep_s4}}
    \phantomsubfloat{\label{fig:AI_sep_s4}}
    \phantomsubfloat{\label{fig:ecpt_s4}}
    \phantomsubfloat{\label{fig:ecpt_s4_2}}
    \phantomsubfloat{\label{fig:ecpt_s4_3}}
    \vspace{-2\baselineskip}
    \caption{Key observables for the fourth state. 
        \textbf{(a):} interatomic separation distribution $\rho^{AA}_1(z_{A} - z_{A}^{\prime})$ for different atom-atom coupling strengths $g$. The inset shows the expectation value $\braket{d_{AA}}$ for the atom-atom separation as a function of $g$ (eq.~\eqref{eq:d_AA}).  \textbf{(b):} interspecies separation distribution $\rho^{AI}_1(z_{A} - z_{I})$ for varying atom-atom coupling strengths $g$. The inset shows the expectation values $\braket{d_{AI}}$ for the atom-ion separation as a function of $g$ (eq.~\eqref{eq:d_AI}). \textbf{(c)-(e):} the evolution of the laboratory frame energy components with atom-atom coupling strength $g$. \textit{Note for (a) and (b):} the grey lines indicate the distance between the minima of the atom-ion interaction potential \eqref{eq:atom-ion-int}. Due to the parity symmetry it is sufficient to show only the positive semi-axis. \textit{All subfigures:} the solid curves correspond to a mobile ion, whilst dashed curves correspond to a static ion.
    }
     \label{fig:s4_observables}   
\end{figure*}
\begin{figure*}
    \centering
    \includegraphics[width=1.8\columnwidth]{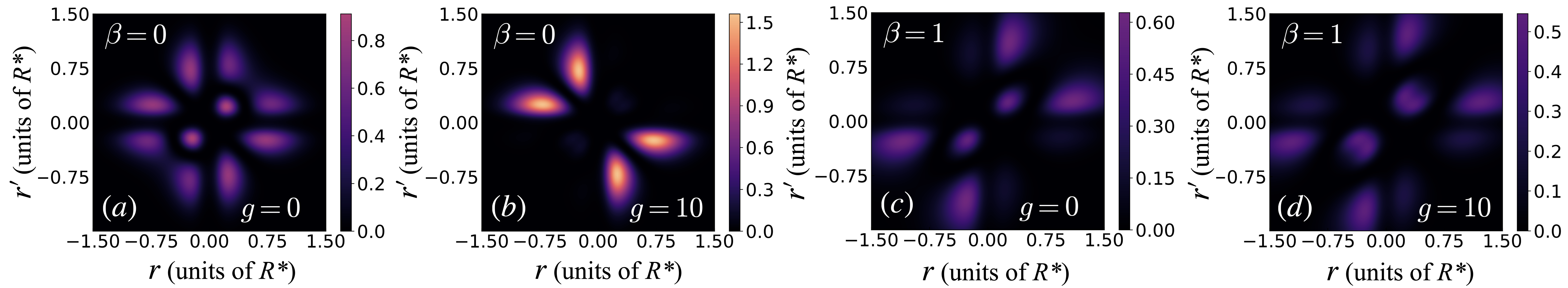}
    \phantomsubfloat{\label{fig:dmat2_s4_1}}
    \phantomsubfloat{\label{fig:dmat2_s4_2}}
    \phantomsubfloat{\label{fig:dmat2_s4_3}}
    \phantomsubfloat{\label{fig:dmat2_s4_4}}
    \vspace{-2\baselineskip}
    \caption{
        Snapshots of the atomic probability density $\rho_2^{\text{AA}} = |\psi(r,r^{\prime})|^2$ of the fourth excited state for different interaction strengths $g$ for a static ion ((a) and (b)) and a mobile ion ((c) and (d)).
    }
    \label{fig:dmat2_s4}
\end{figure*}

\subsubsection{Static ion}~\label{ssec:s4_static}
The fourth excited state for two non-interacting  ($g=0$) atoms 
coupled to a static ($\beta=0$) ion 
is given by the number state $\ket{1,0,0,1}$, with one atom 
in the even molecular orbital 
$\phi_0$ and the other in the odd vibrational orbital 
$\phi_3$ (see~\cref{fig:h1b_solutions}).\\
Similar to the third excited state, the total energy (given in~\cref{fig:spectrum})
and the energy components (see~\cref{fig:ecpt_s4,fig:ecpt_s4_2,fig:ecpt_s4_3})
are robust to $g$ variation. 
The small energy gap between the third and fourth
eigenstates is due to a difference 
in the potential energy $P_A$ of the atoms.
Nevertheless, this state also exhibits structural changes
in the interatomic separation distribution $\rho_1^{AA}$ 
(dashed curves in~\cref{fig:sep_s4}).
Contrary to the third excited state, here we observe two humps:
a narrow hump at $r_A=z_A-z'_A=0$ and a broad hump at $r_A \sim 1.0 R^*$.
As $g$ increases, the hump at $r_A=0$ becomes smaller and broader,
whilst the other becomes higher and narrower.
As a result, the distance among the atoms $d_{AA}$ increases with $g$
(black dashed curve in the inset of~\cref{fig:sep_s4}).\\
In the two-body density $\rho_2(r,r')$ at $g=0$ (see~\cref{fig:dmat2_s4_1}),
we observe the absence of anti-bunched regions 
in close proximity to the ion ($r<0.3R^* ,r'<0.3 R^*$).
This is in contrast to the third excited state and explains the absence
of the third peak at $r_A=0.4 R^*$.
Additionally, since the occupied orbitals $\phi_0$ 
and $\phi_3$ are of opposite parity symmetry, 
the off-diagonal at $r=-r'$ is zero.
At stronger coupling $g=10$ (see~\cref{fig:dmat2_s4_2}) the only eligible 
configurations are the ones, 
where one atom is bound to ion and the other 
is unbound on the opposite side
w.r.t.\ the ion.
\subsubsection{Mobile ion}~\label{ssec:s4_mobile}
For $\beta=1$, the total energy and energy components are not as robust to variations in $g$. Notably, there is a rapid increase in $V_{AA}$ given by $g\rho_1^{AA}(r=0)$ (solid light blue curve in~\cref{fig:ecpt_s4_3}), since the probability amplitude at $\rho_1^{AA}(r=0)$ is unaffected by $g$ (see solid curves in~\cref{fig:sep_s4}). As a result, the energy gap between the third and fourth excited state widens with growing $g$ (see~\cref{fig:spectrum}).

\section{Summary and Outlook}\label{sec:conclusion}
In this study, we have analysed the low-energy eigenstates
of an atom-ion hybrid system consisting of a pair of bosons interacting with a single ion, 
where both species are confined in a quasi-1D trapping geometry. 
The eigenstates were obtained by means 
of the Multi-Layer Multi-Configuration Time-Dependent Hartree method for Bosons (ML-MCTDHB), 
an \textit{ab initio} method for simulating 
entangled mixtures 
with significant intra-component correlations.
We described the eigenenergies' dependence on the
atom-atom coupling strength $g$ 
for the case of an infinitely-heavy ion ($\beta=0$) and contrasted this to the case of an equal mass system ($\beta=1$).
The former we termed the \textit{static} ion system 
and the latter the \textit{mobile} ion system.
Each eigenstate has been characterised in terms
of its interatomic and interspecies separation distributions, 
average separations among the particles, 
and energy allocation in different Hamiltonian components.\\
In general, the repulsive interaction between the atoms increases their average separation,
accompanied by a broadening of the interatomic separation distribution and an
energy exchange between atomic kinetic and atom-ion interaction energies. 
The average distance to the ion does not necessarily change in the process, however. When it does change, the atoms separate, whilst simultaneously remaining on the same side w.r.t the ion
(first and second excited states). On the other hand, a constant atom-ion distance indicates that the atoms have transitioned to a configuration in which the ion lies in-between them. In this case, both atoms are either bound (ground state) or one atom moves freely in the harmonic trap 
(third and fourth excited states).
Contrary to the repulsive interatomic interaction, the mobility of the ion works to increase the average separation between all particles, irrespective of the species.\\
We explained the apparent ion mobility-induced bunching effect observed in the ground state through use of an effective potential, which predicts that the ion mobility 
morphs the standard double-well-like potential produced by the static ion into a pseudo-harmonic potential, 
such that the atoms cluster together at the trap centre. Likewise, the effective potential for the ion predicts that for strong interatomic interactions, the anti-bunched atom pair acts like a pincer, which confines the ion at the centre of the trap. These predictions agree with the trend in the ion's energy components obtained via exact numerical methods.\\
Regarding possible experimental realisation of our hybrid model, 
it is important to note that we have neglected all three-body recombination processes.
However, considering the low particle density in few-body systems such as ours,
the loss rate is expected to be insignificant.
For larger particle numbers in one spatial dimension, such decay channels 
may be suppressed by strong repulsion 
among the atoms \cite{Gangardt2003}. 
We have additionally neglected charge transfer and radiative loss
resulting from reactions between the ion and the atoms.
For certain heteronuclear atom-ion pairings,  
these inelastic processes happen to be of low probability~\cite{Tomza2015}, 
and for other pairings the chemical reactivity can be controlled 
through the use of a magnetic field~\cite{tomza2015cold}. 
We have further assumed our system is at temperatures low enough for pure s-wave scattering, which has long been the goal of atom-ion experiments. Recent experiments have attained temperatures at the threshold of this regime via buffer gas cooling of a single ion in a Paul trap for species pairings with a large mass-imbalance~\cite{feldker2020buffer}, corresponding to a static ion system. 
As for the mobile ion system, 
the recent advent of optical traps for ions~\cite{schaetz2017trapping} 
provides a promising platform that could be used for experiments with atoms and ions of the same element.\\
This work has mapped out the landscape of stationary states 
for an exemplary few-body mixture characterised by long-range interspecies interactions. It lays the foundation for the route to more exotic and complex systems, such as 
solid-state emulation in Coulomb crystals and dipolar quantum gases. Additionally, exploring how the properties of the present atom-ion hybrid system evolve with greater particle numbers, alternative species pairings, and differing trapping frequencies ($\eta \neq 1$) would be an interesting avenue for future study.
This work also focused solely on a time-independent problem,
therefore a natural extension would be to consider many-body dynamics,
e.g.: ion immersion in an atomic gas and time-resolved monitoring of atom capture,
leading to the formation of mesoscopic molecules. Other theoretical investigations into atom-ion hybrid systems have employed a variety of techniques, including: exact diagonalisation~\cite{garcia-march2014distinguish}, quantum Monte Carlo~\cite{garcia-march2013sharp,astrakharchik2020ionic}, density matrix renormalisation group (DMRG)~\cite{michelsen2019}, variational methods~\cite{rath2013field}, semi-classical methods~\cite{melezhik2019} and hyperspherical coordinates~\cite{perez2018universal}, for which our work may serve as a useful numerical benchmark.\\

\section*{Acknowledgements}
D. J. B. thanks Kevin Keiler and Fabian K\"ohler for helpful discussions regarding the numerical implementation. M. P. gratefully acknowledges a scholarship of the Studienstiftung des deutschen Volkes. This work is funded by the Cluster of Excellence: 'Advanced Imaging of Matter' of the Deutsche Forschungsgemeinschaft (DFG) – EXC 2056 – project ID 390715994.
\bibliographystyle{apsrev4-1}
\bibliography{references}
\end{document}